\documentclass[lettersize,journal]{IEEEtran}
\usepackage{amsmath,amsfonts}
\usepackage{amssymb}
\usepackage{algorithmic}
\usepackage{algorithm}
\usepackage{array}
\usepackage[caption=false,font=normalsize,labelfont=sf,textfont=sf]{subfig}
\usepackage{textcomp}
\usepackage{stfloats}
\usepackage{url}
\usepackage{verbatim}
\usepackage{graphicx}
\usepackage{cite}
\begin{document}

\onecolumn
\null
\thispagestyle{empty}
\begin{center}
This work has been submitted to the IEEE for possible publication. 
Copyright may be transferred without notice, after which this version may no longer be accessible.
\end{center}
\newpage
\setcounter{page}{1}
\twocolumn

\title{System-Level Analysis of Module Uncertainty Quantification in the Autonomy Pipeline}

\author{Sampada Deglurkar*, Haotian Shen, Anish Muthali, 
Marco Pavone,~\IEEEmembership{Member,~IEEE}, Dragos Margineantu, ~\IEEEmembership{Member,~IEEE}, Peter Karkus, Boris Ivanovic, Claire J. Tomlin,~\IEEEmembership{Fellow,~IEEE}

\thanks{S. Deglurkar and C. J. Tomlin are with the Department of Electrical Engineering and Computer Sciences, University of California, Berkeley. Work by H. Shen and A. Muthali was done here.
        {\tt\small \{sampada\_deglurkar, tomlin, dalyshen, anishmuthali\}@berkeley.edu}}%
\thanks{M. Pavone, P. Karkus, and B. Ivanovic are with NVIDIA Research and M. Pavone is also with the Department of Aeronautics and Astronautics, Stanford University
        {\tt\small \{mpavone, pkarkus, bivanovic\}@nvidia.com}}
\thanks{D. Margineantu is with The Boeing Company
        {\tt\small dragos.d.margineantu @boeing.com}}
\thanks{*Corresponding author \\
This work is an extension of \cite{deglurkar2024}, which was presented at CDC 2024.}
}

\markboth{IEEE Transactions on Control Systems Technology, 2026}%
{Shell \MakeLowercase{\textit{et al.}}: }

 \maketitle

\begin{abstract}

Modern autonomous systems with machine learning components often use uncertainty quantification to help produce assurances about system operation.
However, there is a lack of consensus in the community on what uncertainty is and how to perform uncertainty quantification.
In this work, we propose that uncertainty measures should be understood within the context of overall system design and operation.
To this end, we present two novel analysis techniques.
First, we produce a probabilistic specification on a module's uncertainty measure given a system specification.
Second, we propose a method to measure a system's input-output robustness in order to compare system designs and quantify the impact of making a system uncertainty-aware.
In addition to this theoretical work, we present the application of these analyses on two real-world autonomous systems: an autonomous driving system and an aircraft runway incursion detection system.
We show that our analyses can determine desired relationships between module uncertainty and error, provide visualizations of how well an uncertainty measure is being used by a system, produce principled comparisons between different uncertainty measures and decision-making algorithm designs, and provide insights into system vulnerabilities and tradeoffs.

\end{abstract}


\section{Introduction}
As autonomous systems become more pervasive in industry and in daily life, we see simultaneously the trends towards incorporating machine learning into these systems to give them greater generality and flexibility as well as the need for these systems to be safe and reliable.
Many of these systems are modularized for interpretability and ease of engineering design.
For example, a classic robotics system architecture may involve perception and localization modules, whose outputs are then passed on to planning and control modules.
Some modules, such as the perception module, may involve learning, while others, such as the control module, may not.
In recent times, producing assurances on system operation in the presence of these machine learning components has been an especial challenge.

\begin{figure}[!t]
\centering
\includegraphics[width=3.53in]{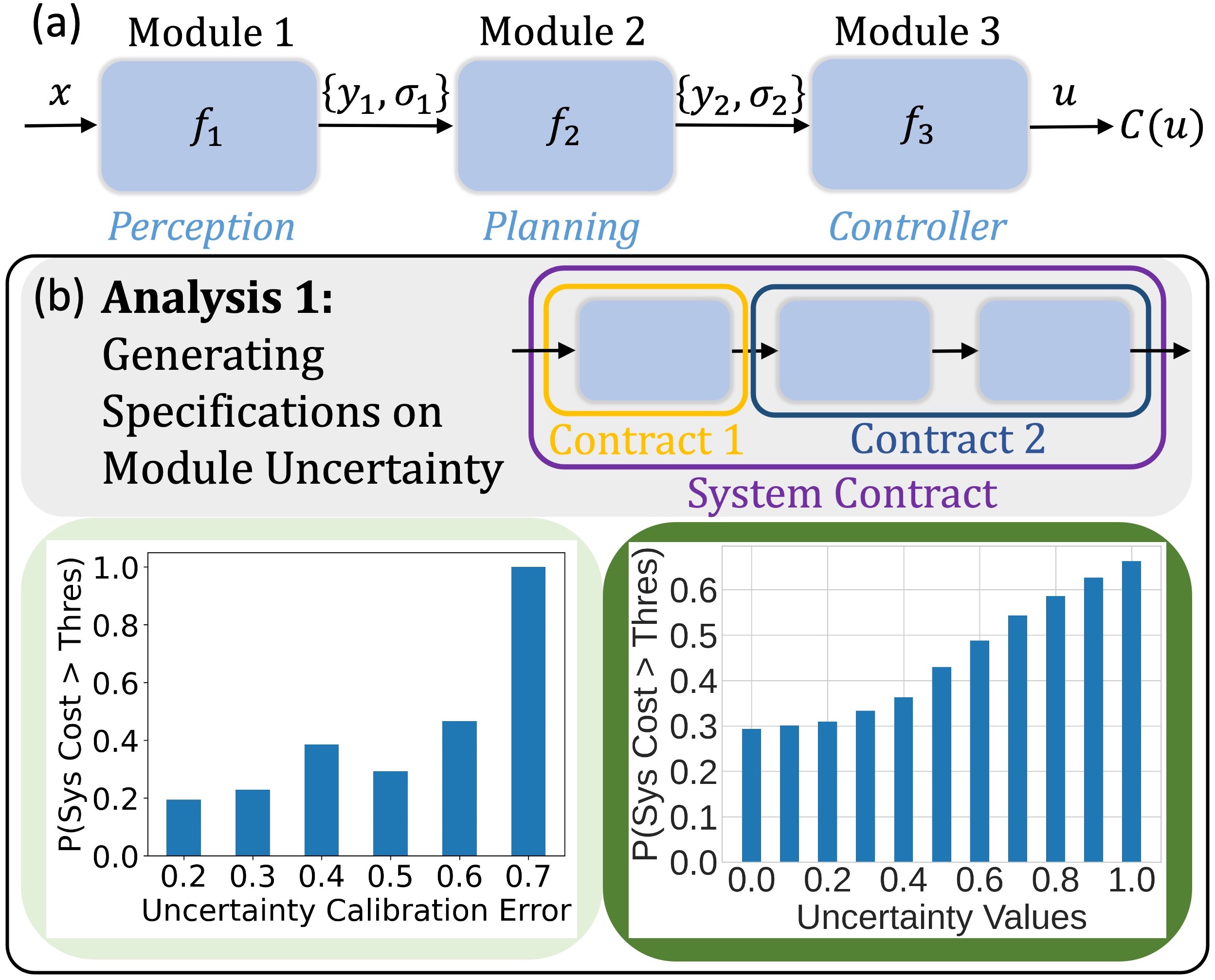}
\centering
\includegraphics[width=3.53in]{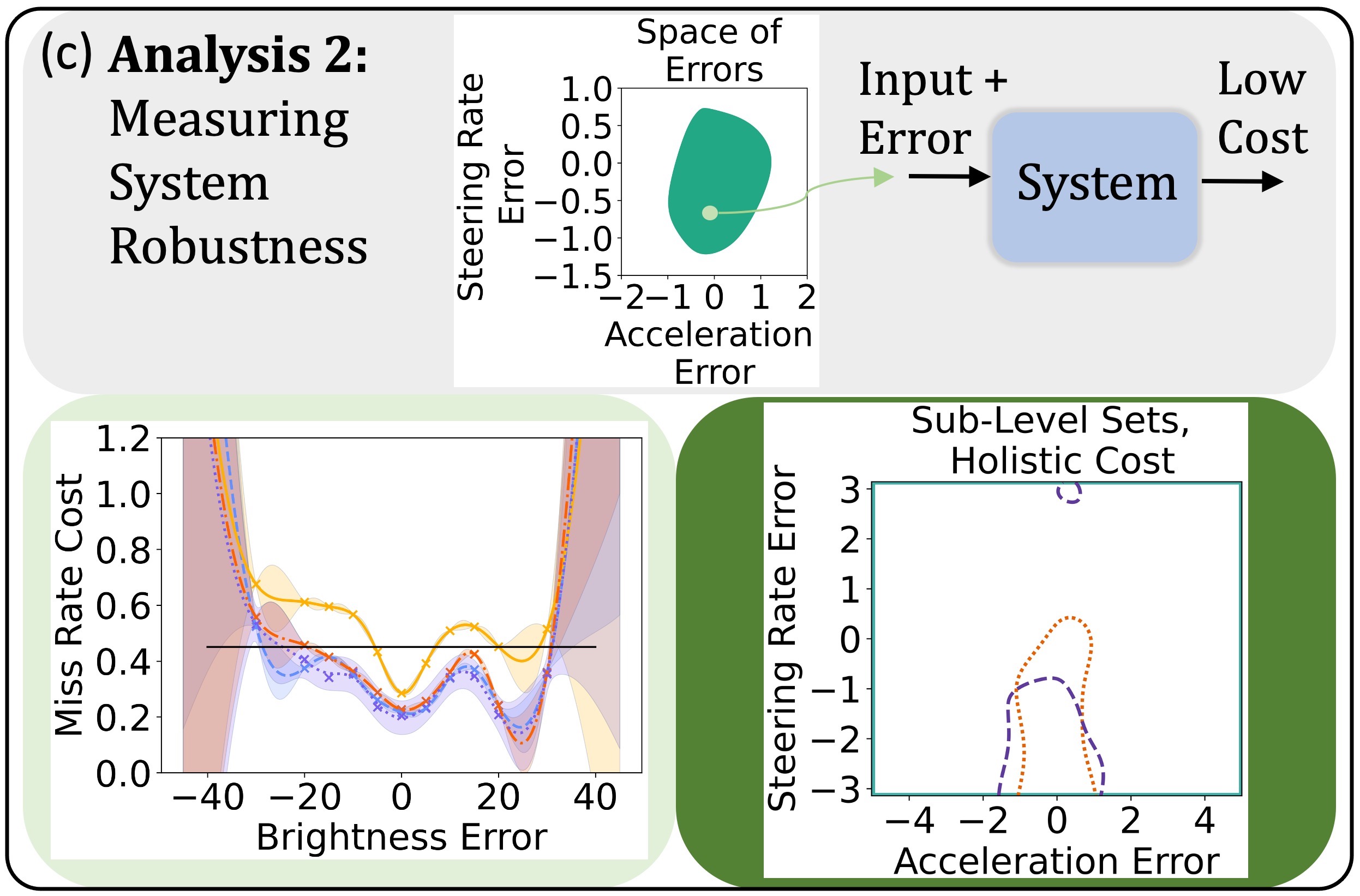}
\caption{
(a) An example of a sequential, modularized autonomy pipeline, in which $x$ is the state, $u$ is the control input, and $C(u)$ represents a cost function.
Module $i$ has output $y_i$ with associated uncertainty $\sigma_i$.
(b) Analysis 1: We propose that generating a specification on a module's uncertainty measure will help us to apply our system-level perspective. 
Using assume-guarantee contract theory, we treat this as a problem of finding the contract of an upstream module of interest (Contract 1) given the system contract and the contract of the downstream decision-making modules (Contract 2).
We show some example results in which we use data to generate histograms that help to characterize the Contract 2. 
(c) Analysis 2: We propose to measure a system's input-output robustness by searching for all input errors such that the system output incurs an evaluation cost below a threshold.
In the example results shown, we display functions and level sets that are the solution to the search problem.
Different colors and line styles indicate different system designs.
For both analyses, the light green color indicates evaluation on the Runway Incursion Detection system and the dark green color indicates evaluation on the Autonomous Driving system.}
\label{fig:cover_figure}
\end{figure}

In the machine learning literature, a common way to provide more trust in learned models is to perform uncertainty quantification \cite{Cai2023EVORADE, Loquercio2020}.
Techniques such as anomaly detection \cite{basseville_anomaly_detection}, out-of-distribution (OOD) detection \cite{Sharma2021SketchingCF, pmlr-v162-hendrycks22a, liu_2020, srikant2018, ancha2024}, and epistemic and aleatoric uncertainty quantification \cite{kendall_what_uncertainties_2017, pmlr-v48-gal16, ensembles2017, Postels2019SamplingFreeEU, Lee2021TrustYR}, exist to help a model ``know that it does not know''.
However, the uncertainty measures produced by these techniques can often be unintuitive \cite{charpentier2022, yao2019quality} and require calibration \cite{angelopoulos2021gentle, Muthali2023, dixit_lindemann_wei_cleaveland_pappas_burdick_2022}.
Additionally, there is no consensus on how to define some of the terminology for these techniques, such as ``out-of-distribution'', and it is not often clear which technique one should use \cite{Postels2021OnTP}.
Thus, the uncertainty quantification technique and the type of decision-making-under-uncertainty algorithm used by a system tend to be heuristic design choices.

While much of the work in the machine learning literature regarding uncertainty quantification considers standalone modules, in the current paradigm, learned modules are part of a larger decision-making system.
As such, we argue that uncertainty measures are given meaning not by themselves, but in the context of how they are used by the system in the operating environment.
That is, the uncertainty measure of a module should be understood \textit{relative} to the efficacy of the decision-making-under-uncertainty algorithm employed by the downstream modules.

In this paper, we present two analysis techniques to provide a better understanding of uncertainty quantification from a system-level perspective.
The two analyses are united by a desire to provide a theoretical framing for empirically-driven design processes.
An overview of our work is presented in Figure \ref{fig:cover_figure}.
In our first analysis, we study uncertainty quantification as it governs individual connections between modules.
Our second analysis considers the impact of uncertainty-awareness on the overall pipeline.

In our first analysis, we consider the problem of generating a specification on a module's uncertainty measure, given a system-level specification.
We use assume-guarantee contract theory to formally define the problem.
We then write our specification as a particular inequality that connects three quantities: the relationship between a module's uncertainty measure and its error, how that uncertainty measure is used downstream, and the overall system performance.
Our framework allows us to simultaneously identify the characteristics of the uncertainty measure and the downstream decision-maker that will lead to good overall system performance.

In our second analysis, we propose a method to measure a system's robustness, essentially finding the set of all possible input perturbations under which the system, on average, incurs a cost below a threshold.
This becomes a sub-level set estimation problem.
The robustness metric allows us to compare uncertainty-aware and non-uncertainty-aware system designs in terms of both performance and robustness.
Finally, we apply our analyses on two realistic, complex, and industry-grade autonomous systems: an autonomous driving system and an aircraft runway incursion detection system.
Given that system-level design questions such as ours can be computationally intensive to answer, we also include discussion of the computational complexity of our methods and the tradeoffs that can be necessary to make our analyses feasible.

This work is an extension of \cite{deglurkar2024}, which was presented at the Conference on Decision and Control (CDC), 2024.
The main novel additions that are present in this work are as follows:
\begin{itemize}
    \item We provide a deeper development of our specification generation formulation using assume-guarantee contract theory.
    That is, in order to define the system properties that are subject to the specification generation, we use the theory to enumerate criteria that the properties must abide by.
    We also add \textit{environment semantics} as an important part of the properties.
    \item We demonstrate a wider applicability of both analysis techniques. 
    In \cite{deglurkar2024}, we use each analysis technique to analyze only one real-world autonomous system each.
    Here, we apply both analysis techniques to both autonomous systems mentioned above.
    In doing so, we attain more understanding of what our techniques can truly achieve.
\end{itemize}

This paper is organized as follows: Section \ref{sec:related_work} provides an overview of related work.
In Section \ref{sec:defns_assumptions}, we give definitions and assumptions, and in Section \ref{sec:problem_statements}, we mathematically describe the setups of the two analysis techniques.
Section \ref{sec:case_study_system_descriptions} contains descriptions of the autonomous driving system and aircraft runway incursion detection system.
Section \ref{sec:generating_specifications} discusses our first analysis technique regarding the specification generation problem.
Section \ref{sec:measuring_robustness} discusses our second analysis technique regarding the robustness measurement problem.
We conclude with Section \ref{sec:conclusion}.

\section{Related Work}\label{sec:related_work}

\subsection{Modules and Uncertainty Measures in the Context of the System}

The fundamental concept behind our work is that of taking a system-level perspective on the characteristics of individual modules.
This philosophy is drawn from several recent works.
In \cite{ivanovic2021} and \cite{Nakamura2024NotAE}, module errors are considered in terms of their downstream impact.
Doing so can help to re-design modules in a system-level risk-aware way, such as in  \cite{Corso2022RiskDrivenDO}.
Works such as \cite{Zhang2024WhyCY} and 
\cite{matni_coordinating_planning_tracking} also design certain modules in an autonomy pipeline by explicitly taking into account how they interact with other modules.

Various works also connect module uncertainty measures with system-level objectives.
\cite{azizan2022} and \cite{pmlr-v164-farid22a} aim to produce an uncertainty measure and OOD detector, respectively, according to the desideratum that they should help to predict downstream performance.
In contrast, we try to shift away from defining a new or better uncertainty measure and towards a deeper understanding of the role of uncertainty quantification in the overall system.
In the first analysis in our work, we use assume-guarantee contract theory to define what \textit{property} we care about with respect to a module's uncertainty measure.
Thus, we essentially define desiderata for an uncertainty measure in a more rigorous way.

Similar to our work, several other works connect the concept of \textit{calibration} with system-level perspectives on uncertainty measures.
For example, \cite{lekeufack2024decision} calibrates the decision outputs of the autonomy pipeline directly with respect to risk instead of calibrating upstream uncertainty measures. 
The works \cite{yeh2024endtoend} and \cite{kiyani2025optimaluncertainty} also recognize the need for calibrated prediction sets produced at the module level.
Both works ask how to produce prediction sets representing a learned model's uncertainty that are \textit{optimal} with respect to a downstream decision-making cost, for example using end-to-end learning. 

In our work, we present a more general perspective, in which we treat \textit{any} intelligent downstream decision-maker as performing calibration of the upstream uncertainty measure with respect to system-level costs, which is what gives meaning to an uncertainty measure in the first place.
Indeed, we can quantitatively understand a decision-maker's efficacy as the degree to which this occurs.
Additionally, we allow for more heuristic approaches to uncertainty representation and decision-making, as can be more commonly implemented in real-world systems.
While works such as \cite{kiyani2025optimaluncertainty} can make theoretical statements about optimality, they only consider uncertainty representations in the form of prediction sets and decision-making in the form of robust optimization.
Thus, our work has a more practical orientation.

\subsection{Specification Generation and Assume-Guarantee Contract Theory}

In our first analysis technique, we aim to produce a module-level specification given a system-level specification.
One perspective on this problem is given by Katz et al \cite{katz2023}, in which the system-level specification is that the probability of a system-level failure should be below a threshold.
Finding all module behaviors such that this is true is represented and solved as a sub-level set estimation problem, with all desired module behaviors therefore lying within the set.

One can also turn to the formal methods literature more broadly to consider the specification generation problem.
In this work, we use assume-guarantee contract theory \cite{SangiovanniVincentelliDesignForCyberPhysical, nuzzo_contracts, Incer:EECS-2022-99}, which allows us to formally define module properties and interconnections. 
In particular, we draw upon the concept of the quotient contract as introduced by Incer et al \cite{incer_quotients}, \cite{Incer2023Pacti}.
We extend \cite{Incer2023Pacti} by producing a novel quotient contract formulation involving module uncertainty measures.
Our formulation is additionally novel due to its particular \textit{probabilistic} form.

\subsection{Understanding System Robustness via Input Perturbations}

In our second analysis technique, we aim to design a robustness metric for evaluating systems, essentially serving as a system design objective.
In producing a new design objective to \textit{justify} why uncertainty quantification is useful, we take inspiration from \cite{Matni2024}, which derives the classic layered architecture in control systems from an optimization problem.
Our robustness metric involves a search over perturbations or errors to the input to the system such that the system output has low cost.
This problem is similar to that explored by \cite{Akella2021}, and the tools we use for the level-set estimation that this problem entails are also explored by prior works such as \cite{katz2023, Iwazaki2020, Inatsu2021} and \cite{Gotovos2013}. 
Our work differs from these as follows: (1) We explicitly use our sub-level set to produce a \textit{measurement} of robustness, and connect that measurement with choices of uncertainty-aware system designs, and (2) We examine how different system designs trade off between performance and this robustness metric, thus connecting uncertainty quantification with system design objectives.

\section{Definitions and Assumptions}\label{sec:defns_assumptions}

Our key focus is on analyzing modularized autonomy pipelines involving components connected in sequence.
Such pipelines are usually designed to first involve estimation-related modules, such as those performing perception or filtering, followed by decision-making-related modules, such as those performing high-level planning and low-level tracking and control.
We define ``system'' as this modularized pipeline, as opposed to the controlled object that uses the controls to evolve the state.

Mathematically, we write each module as a function $f_i$, where $i \in \{1,...,n\}$ indexes the module for $n$ modules.
The input to the system is the state $x \in \mathcal{X}$, which is a random variable drawn from a distribution $\mathcal{D}$.
The output of the system is the control command $u \in \mathcal{U}$.
Each module produces a nominal output $y_i \in \mathcal{Y}_i$ and also may be associated with an uncertainty quantification technique that produces an uncertainty measure $\sigma_i \in \Xi_i$.
Additionally, we can compute a measure of error for each module given $y_i$ and a ground truth value $y_i^* \in \mathcal{Y}_i$.
We denote module $i$'s error value as $e_i \in \mathcal{E}_i$.
Finally, we have access to an evaluation cost function $C: \mathcal{U} \to \mathbb{R}$.
We point out that since $x$ is a random variable, every $y_i$, $e_i$, and $\sigma_i$ is as well, along with $u$ and $C(u)$.
Though we place no restrictions on $\mathcal{X}$, $\mathcal{U}$, and all $\mathcal{Y}_i, \mathcal{E}_i$, and $\Xi_i$, for many systems these are often spaces of vectors, matrices, distributions, or sets.
For example, each $\sigma_i$ may be a vector, distribution, or simply a scalar. 

Let $\Omega_i$ be the space of sets of the form $\{a, b\}, a \in \mathcal{Y}_i, b \in \Xi_i$ for $i \in \{1,...,n\}$.
Formally, we write that $\mathcal{U} = \mathcal{Y}_n$ and that $f_1: \mathcal{X} \to \Omega_i$ and $f_i: \Omega_{i-1} \to \Omega_i \; \forall i \in \{2,...,n\}$.
Given a module $f_{i'}$, we denote the ``decision-making-under-uncertainty algorithm'' relative to this module as the function composition $f_n \circ ... \circ f_{i' + 1}$.
In the case that a module does not involve uncertainty quantification, the associated $\sigma$ simply takes on a `null' value.

We will also find it necessary to define the discrete random variable $z$ that represents the type of environmental scenario that the system encounters.
Since $z$ is discrete and environmental information is often present in $x$, $z$ can be thought of as a compression of $x$. 

A key assumption in this work is that each module is deterministic and time-invariant given these input-output definitions.
This means that each forward pass of the pipeline corresponds to an open-loop decision-making output.
As discussed in our Case Studies, this assumption may require definitions of $x$, $u$, and $C(u)$  that capture all the information needed to run the system and evaluate its performance.
For example, $x$ can involve a history of observations and $u$ can be a sequence of actions.

\section{Problem Statements}\label{sec:problem_statements}
Our analyses aim to answer two questions about modular systems: (1) How can we generate requirements on the uncertainty measure of a module given system-level requirements?, and (2) How can we evaluate the impact of uncertainty quantification on the robustness of a system? 
The first question examines interconnections between modules that are producing and using uncertainty measures, and the second helps a designer to choose between alternative system designs for an application.

\subsection{Generating Specifications on Uncertainty Measures}\label{sec:problem_generating_specs}

As we consider contextualizing uncertainty measures within the overall system, we require a way to reason about how the value of an uncertainty measure can lead to either desirable or undesirable system outcomes.
For this reason, for our first analysis technique we choose to study the question of how to generate a module-level specification on uncertainty measure values given a system-level specification.
In our setting, we consider a \textit{probabilistic} system-level specification:

\begin{equation}\label{eqn:system_spec}
    P(C(u) > c) < \alpha
\end{equation}
for designer-chosen thresholds $c$ and $\alpha$.
This specification thus describes the set of allowable distributions over $C(u)$.

We would like to produce a specification to constrain the values of a target module's uncertainty measure.
However, this problem as written is not clearly defined. 
In the same way that the system specification is a constraint on a distribution, what property should the module-level specification constrain?
Here, a \textit{property} is defined as a set of behaviors \cite{Incer:EECS-2022-99}.
However, an uncertainty measure by itself is not a module property but the output of an uncertainty quantification technique, and has many characteristics that we may want to constrain.

We turn to assume-guarantee contract theory to define our properties.
An assume-guarantee contract $(A, G)$ is a pair of properties, the \textit{assumptions} $A$ and \textit{guarantees} $G$.
A module or a system implements a contract if it satisfies the guarantees given that it operates in any environment that satisfies the assumptions \cite{Incer:EECS-2022-99}.
The problem of obtaining a module-level specification given a system-level specification is very related to that of taking a \textit{quotient contract} \cite{incer_quotients}.
That is, given the system-level contract and the contract of the remaining modules of the system, one can compute the contract of the module of interest.
If the contract of the remaining system modules is not already known, it must be obtained, for example using data-driven characterizations.
In Section \ref{sec:generating_specifications}, we use this framework to write down specific criteria that a designer must consider when defining uncertainty-related properties.
We then proceed to solve a quotient contract problem.

\subsection{Measuring System Robustness}\label{sec:problem_robustness}

For our second analysis, we desire to understand uncertainty-aware systems by examining their placement in the broader system design optimization landscape.
Performance objectives often form an axis in this multi-objective space.
In this work, we argue that uncertainty-awareness arises from considering a \textit{robustness} objective in addition to that of performance.
We thus propose an input-output robustness metric that can form this second objective.

In this problem statement, we define the action of the system as $f := f_n \circ f_{n-1} \circ... \circ f_1$.
A robust system is one that is performant even under perturbations to the input distribution $\mathcal{D}$. In our definition, this occurs when each $x$ sampled from $\mathcal{D}$ is shifted to $p_{\epsilon}(x)$, where $p$ is a known function with parameter $\epsilon$.
This shift or perturbation can arise from sensor errors, adversarial interference, or a disturbance in the environment.
If $x$ belongs to a high-dimensional space, $p_{\epsilon}(x)$ can represent a change to any of its semantic attributes.
We thus formalize the notion of good performance under an input perturbation as the condition

\begin{equation}\label{eqn:robustness_one_eps}
    \mathbb{E}_{x \sim \mathcal{D}} [(C \circ f)(p_{\epsilon}(x))] < c
\end{equation}
for a designer-chosen threshold $c$.
We briefly note here that although our formulation involves an expected cost value, other statistics can also be considered.
Doing so may require a different methodology for calculating the final robustness metric, but would contain the same ideas that we describe.

We would now like to search for \textit{all} possible $\epsilon$ values such that the condition (\ref{eqn:robustness_one_eps}) is satisfied.
Intuitively, if a system satisfies the condition for all $\epsilon$ values in a set with larger volume, it is performant under more distribution shift ``amounts'' and so is more robust.
Formally, we need to find the set $\mathcal{S}$, where

\begin{equation} \label{eqn:robustness_set}
    \mathcal{S} := \{\epsilon \mid \mathbb{E}_{x \sim \mathcal{D}} [(C \circ f)(p_{\epsilon}(x))] < c \}
\end{equation}
Since a larger set corresponds to a more robust system, our final robustness metric is therefore the size of $\mathcal{S}$.
In Section \ref{sec:measuring_robustness}, we describe how we compute this quantity.

\section{Case Study System Descriptions}\label{sec:case_study_system_descriptions}

\begin{figure}[!t]
\centering
\includegraphics[width=3.3in]{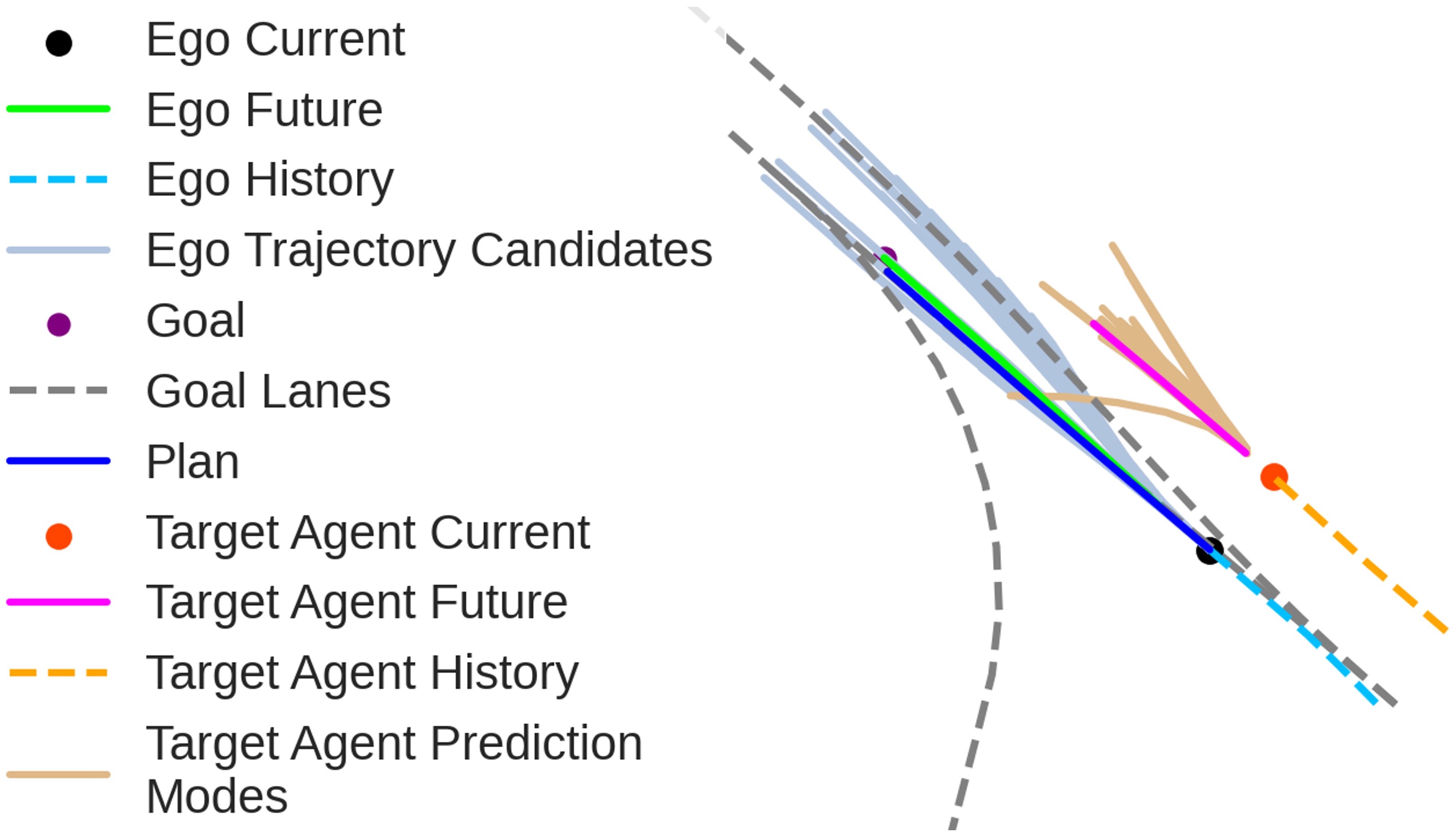}
\caption{The Autonomous Driving system, with an example scene from the nuScenes dataset.
The ego agent must navigate to its goal in the presence of other agents, of which only the nearest agent is shown.
}
\label{fig:case_studies1}
\end{figure}

\subsection{Autonomous Driving System}\label{sec:av_system_description}

One system that we analyze in this work is an autonomous driving software pipeline introduced in \cite{karkus2022diffstack}.
We use the nuScenes dataset \cite{CaesarBankitiEtAl2019}, which contains real driving data from Boston and Singapore.
This system is represented in Figure \ref{fig:case_studies1}.

Similar to Figure \ref{fig:cover_figure}(a), this is a 3-module system comprised of a trajectory forecaster for other agents in the environment ($f_1$), a sampling-based planner ($f_2$), and a controller using MPC ($f_3$).
Here, the trajectory forecaster is the Trajectron++ model \cite{salzmann2020trajectron++}, a recurrent model designed for autonomous driving which takes, among other inputs, non-ego agent state histories and outputs a Gaussian Mixture Model (GMM) representing the distribution of a non-ego agent's future control actions, which are converted into trajectories given the agent's dynamics.
This distribution can help to capture uncertainty heuristics about the non-ego agent's future motion.
For example, we may take the uncertainty measure of the predictor to be (i) the entropy of the most-likely Gaussian of the GMM, which we denote as $\sigma_{ML}$, or (ii) the entropy of the categorical distribution over the GMM's modes, which we write as $\sigma_K$.
These values are normalized to lie in $[0, 1]$.

The planner takes candidate trajectory samples and outputs the trajectory sample that minimizes an internal cost function $C^{\text{int}}$, as seen in \cite{karkus2022diffstack}:

\begin{multline}
    \label{eq:internal_c}
    C^{\text{int}}(\cdot; w) = w_1C_{\text{coll}}(s, \hat{s}) + w_2C_{\text{goal}}(s, g) + w_3C_{l \perp}(s, m) + \\
    w_4C_{l \measuredangle}(s, m) + w_5C_{u}(s)
\end{multline}
where $s$ represents the trajectory sample, $\hat{s}$ represents the trajectory modes of the GMM prediction distribution along with their probabilities, $g$ is the goal, and $m$ is the lane graph.
Each term denotes the cost of collision, the distance to the goal, the lane lateral deviation, the lane heading deviation, and the control effort, respectively.
The $w$ values are weights.
Finally, the MPC controller outputs the final ego agent trajectory by solving an iterative LQR problem optimizing this internal cost, using the planned trajectory as a starting point for the optimization.
Following \cite{karkus2022diffstack}, we evaluate the predictions only for the non-ego agent that is closest to the ego agent at the current time, which we will call the \textit{target} agent.

Thus, we define $x$ to be the history of the target agent for prediction as well as the overall environmental scenario, for example relating to agent relative configurations or road types.
The lengths of the target agent's history and prediction horizon are the same as those in the Trajectron++ work.
We have that $y_1$ is the $\hat{s}$ in Equation (\ref{eq:internal_c}) and $\sigma_1$ is the uncertainty measure that is extracted from the GMM.
The trajectory selected by the planner from the candidates is $y_2$, and the final trajectory optimized by the controller is $y_3$.
We choose two different cost functions $C(u)$ to evaluate the overall system. The cost described in Equation (\ref{eq:internal_c}) can serve not only as a planner internal cost but also as a way to assess output trajectories by their holistic driving behavior.
We also compute a second cost function, the safety cost, as the negative of the minimum distance between the ego agent and all other agents, across all steps in the planning time horizon.

We will consider three different planner designs.
In one design, we will not incorporate either of the uncertainty measures that we have defined; this is the baseline planner as described above.
For our custom uncertainty-aware planner designs, we modify the planner's internal cost function to produce behaviors that are commonly desired in practice.
In our agent-avoiding planner design, the planner reacts to higher uncertainty by becoming more conservative around the target agent: 
\begin{multline} \label{eq:cint_avoid}
    C_{\text{avoid}}^{\text{int}}(\cdot; w) = w_1e^{\alpha \sigma_1} C_{\text{coll}}(s, \hat{s}) + w_2C_{\text{goal}}(s, g) + \\
    w_3C_{l \perp}(s, m) + w_4C_{l \measuredangle}(s, m) + w_5C_{u}(s) 
\end{multline}
with the additional parameter $\alpha$.
In our lane-keeping design, the planner gives less trust to the quality of the collision cost, which uses $\hat{s}$, and prioritizes staying centered in the lane, in response to higher uncertainty.
Lane-keeping is often considered the default safe driving behavior:
\begin{multline}
    \label{eq:cint_lane}
    C_{\text{lane}}^{\text{int}}(\cdot; w) = \sigma_1 (w_1C_{\text{coll}}(s, \hat{s}) + w_2C_{\text{goal}}(s, g)) + \\ \frac{1}{1 - \sigma_1} ( w_3C_{l \perp}(s, m) + w_4C_{l \measuredangle}(s, m) + w_5C_{u}(s))
\end{multline}

\begin{figure}[!t]
\centering
\includegraphics[width=1.85in]{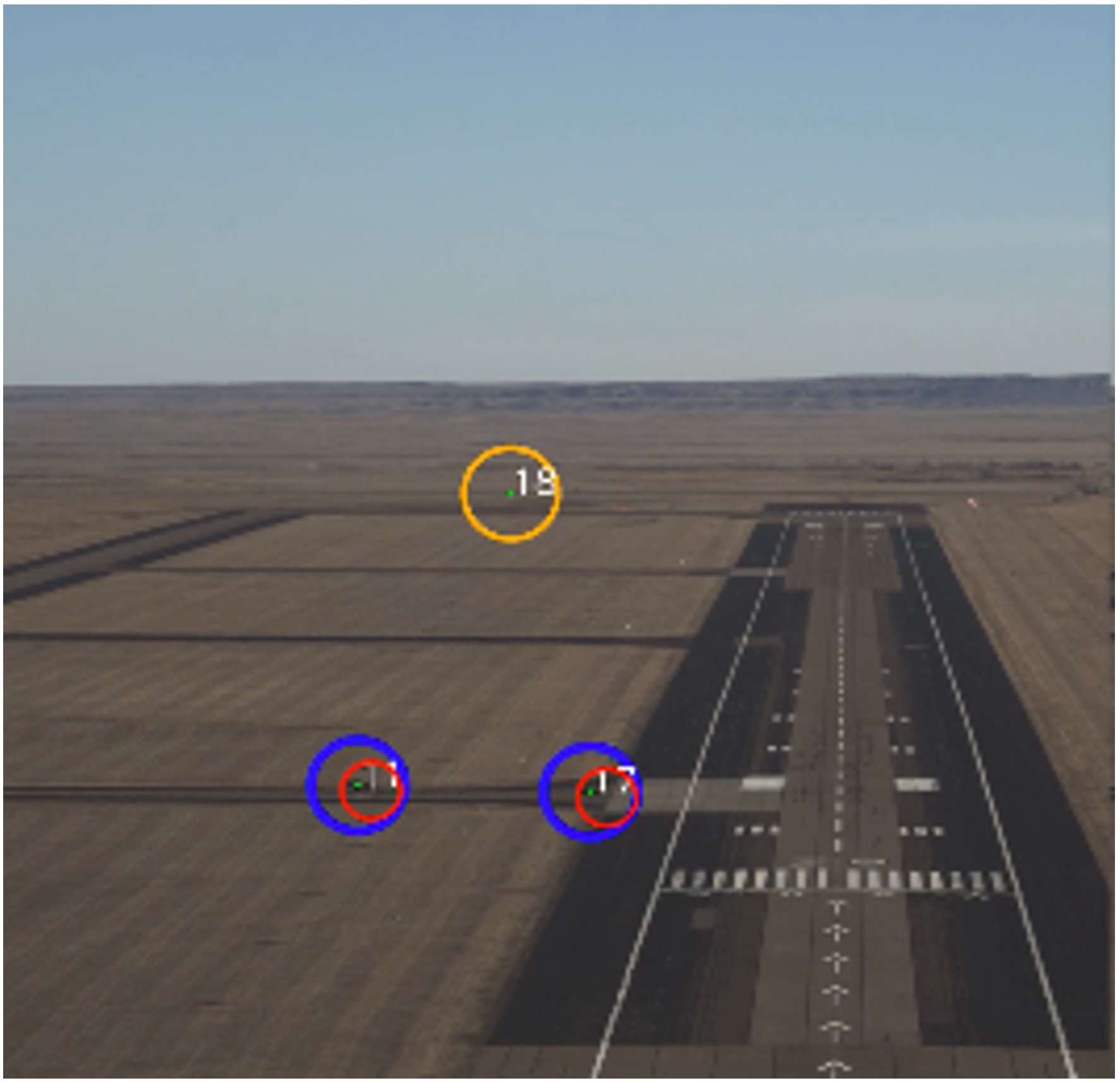}
\caption{The Runway Incursion Detection system, with an example image that might be input to the detector on the aircraft.
From this aerial view of the runway, the aircraft must determine if there exist ground vehicles which could cause a runway incursion.
Red circles correspond to detections.
Yellow circles correspond to ``probationary'' tracks that are not fully initiated yet.
Blue circles correspond to fully initiated tracks.
The numbers correspond to detection confidence scores at that instant in time.
}
\label{fig:case_studies2}
\end{figure}

\subsection{Runway Incursion Detection System}\label{sec:boeing_system_description}

Another real-world system that we analyze is a runway incursion detection system for an aircraft that is about to land.
The goal of the system is to produce advisory information to a pilot to help determine whether to land the aircraft or not, depending on whether an object on the ground, such as a vehicle, is likely to be intruding on the runway.
We study this system with the help of a Boeing aircraft equipped with cameras.
We also use a proprietary Boeing dataset that contains different landing scenarios with possible ground vehicle intrusions.
This system is represented in Figure \ref{fig:case_studies2}.

This is a 2-module system comprised of a detector ($f_1$) and a tracker ($f_2$).
Given images of the runway, the detector produces bounding-box detections of potential objects on or near the runway.
Since the images are taken from the air from potentially far away, there can be a high amount of uncertainty about whether the detected objects truly exist or not.
The detector is a fully convolutional one-stage neural network model \cite{Tian2019}.
It is trained to also produce a confidence score between $0$ and $1$ per detection, where a score closer to $1$ indicates a higher confidence that the detected object exists.
We treat these confidence scores as the uncertainty measure from the detector.

The function of the tracker is then to associate objects to their detections over time and track their positions using a Kalman filtering-based approach.
We design the tracker to be uncertainty-aware by incorporating the detector's confidence scores into a computation of the probability of an object's existence.
These probabilities are used to determine when to initiate and terminate a track, under Wald's sequential probability ratio test (SPRT) \cite{BOOK:Wald}.

Similar to the Autonomous Driving system, we define our relevant system quantities using sequences and histories of observations to meet our time-invariance assumption.
$x$ is defined to be the set of all camera images produced over the course of a track's existence as well as the overall environmental scenario.
$y_1$ is the sequence of bounding-box detections corresponding to a single object track, and $\sigma_1$ is the sequence of confidence scores corresponding to the track.
Finally, $y_2$ is the predicted probability of the track's existence as produced by the tracker.
Using this continuous-valued probability as the output of the pipeline, as opposed to binary values indicating, for example, the presence of an incursion or not, allows us to have a more expressive evaluation cost function.
We set this cost as the absolute difference between $y_2$ and the true probability of existence.
Since the true probability of existence is an ill-defined quantity, we estimate it by dividing the number of predicted detections in a track that correspond to a ground-truth detection by the total number of ground-truth detections in the track.

We will consider four different tracker designs.
In one design, we will not incorporate uncertainty measures at all. 
Instead, the track initiation will be performed by simply checking if detections are being received for a high enough percentage of camera frames.
Track termination will be performed by checking if detections have not been received for a certain amount of frames.
We will also consider three uncertainty-aware tracker designs, denoted by ``Naive'', ``Regression'', and ``Histogram''.
These were produced as part of different system design iterations and are differentiated by how they compute the likelihood ratio in the SPRT.
The Naive tracker computes the likelihood ratio as the detector confidence divided by 1 minus the confidence.
The Regression and Histogram trackers, however, more intelligently use the confidence scores by computing the mean and variance of the confidence scores over a rolling window.
The Regression tracker then uses the mean and variance as features in simple autoregressive linear models for computing the numerator and denominator of the likelihood ratio.
The Histogram tracker, meanwhile, uses prior data to construct histograms offline representing the conditional distributions in the likelihood ratio.
Online, it uses the histograms to look up the probability corresponding to the rolling window's mean and variance.

\section{Generating Specifications on Uncertainty Measures} \label{sec:generating_specifications}

Throughout this section, we will use the Autonomous Driving system with the agent-avoiding planner (using the internal cost function given by Eq. (\ref{eq:cint_avoid})) as our running example.

\textit{\textbf{Running Example:} We focus on $\sigma_1$, which is a scalar value extracted from the output of the trajectory predictor module.
The output of the system, $u$, is the final ego trajectory produced by the system, and the system cost function, $C(u)$, encapsulates desired driving behavior, such as holistically good driving or safe driving.
In our problem statement, we desire a way to use the system-level specification (\ref{eqn:system_spec}) to produce a module-level specification relating to the predictor module's uncertainty measure.}

At a high level, we aim to address the problem in Section \ref{sec:problem_generating_specs} by first defining the uncertainty-aware module property involved in our specification, and then writing the specification as a particular inequality.
We do this using tools from assume-guarantee contract theory (Section \ref{subsec:assume_guarantee}).
Our specification involves a probability distribution that characterizes the behavior of the modules downstream of the target module.
We need to estimate the probabilities in that distribution to have a working specification (Section \ref{sec:estimating_specification}).
Then, we discuss how our specification provides insights into the efficacy of the decision-making-under-uncertainty algorithm and the desired relationship between the uncertainty measure and module error (Section \ref{subsec:specifications_discussion}).
We finally perform evaluations on our Case Studies (Section \ref{subsec:specifications_results_case_studies}).

\subsection{Assume-Guarantee Contract Theory for Uncertainty}\label{subsec:assume_guarantee}

As described in Section \ref{sec:problem_generating_specs}, we will use assume-guarantee contract theory to formally define our problem statement.
Let the module of interest be module $i$.
Then we refer to the contract implemented by modules $1$ to $i$ as $(A_1, G_1)$ and the contract implemented by modules $i+1$ to $n$ as $(A_2, G_2)$.
We write the contract implemented by the system as a whole as $(A_{sys}, G_{sys})$.
We also write $A_i \in \mathcal{A}_i$ and $G_i \in \mathcal{G}_i$ for $i \in \{1, 2\}$ and $A_{sys} \in \mathcal{A}_{sys}$, $G_{sys} \in \mathcal{G}_{sys}$, where $\mathcal{A}_{sys}$, $\mathcal{G}_{sys}$, and all of the $\mathcal{A}_i$ and $\mathcal{G}_i$ are property spaces.  
While we may not have a formal expression for $A_{sys}$, it represents the set of conditions under which the system is designed to operate.
$G_{sys}$ is then the set of distributions over $C(u)$ such that the system-level specification (\ref{eqn:system_spec}) is satisfied.
Since the composition of the $n$ modules describes the system's behavior, we can also set $A_1$ equal to $A_{sys}$ and $G_2$ equal to $G_{sys}$. 

\textit{\textbf{Running Example:} Let us refer to the contract implemented by the trajectory predictor as $(A_1, G_1)$ and the contract implemented by the planner and controller combined as $(A_2, G_2)$.
When we take the quotient contract, we use $(A_{sys}, G_{sys})$ and $(A_2, G_2)$ to compute $(A_1, G_1)$. 
The desired module-level specification would be one that determines $G_1$.
However, we do not know yet what $\mathcal{G}_1$ or $\mathcal{A}_2$ are.
}

We can thus reframe our question as trying to determine the property space $\mathcal{G}_1$.
We know that for a module $i$ implementing contract $(A_i, G_i)$ connected to a module $i+1$ implementing contract $(A_{i+1}, G_{i+1})$, it is true that $G_i \supseteq A_{i+1}$ \cite{Zardini:ETH}.
Intuitively, module $i + 1$ cannot ``request'' more than module $i$ can ``provide''.
So, we write that $\mathcal{G}_1$ must abide by the following criteria:

\begin{figure}[!t]
\centering
\includegraphics[width=3.5in]{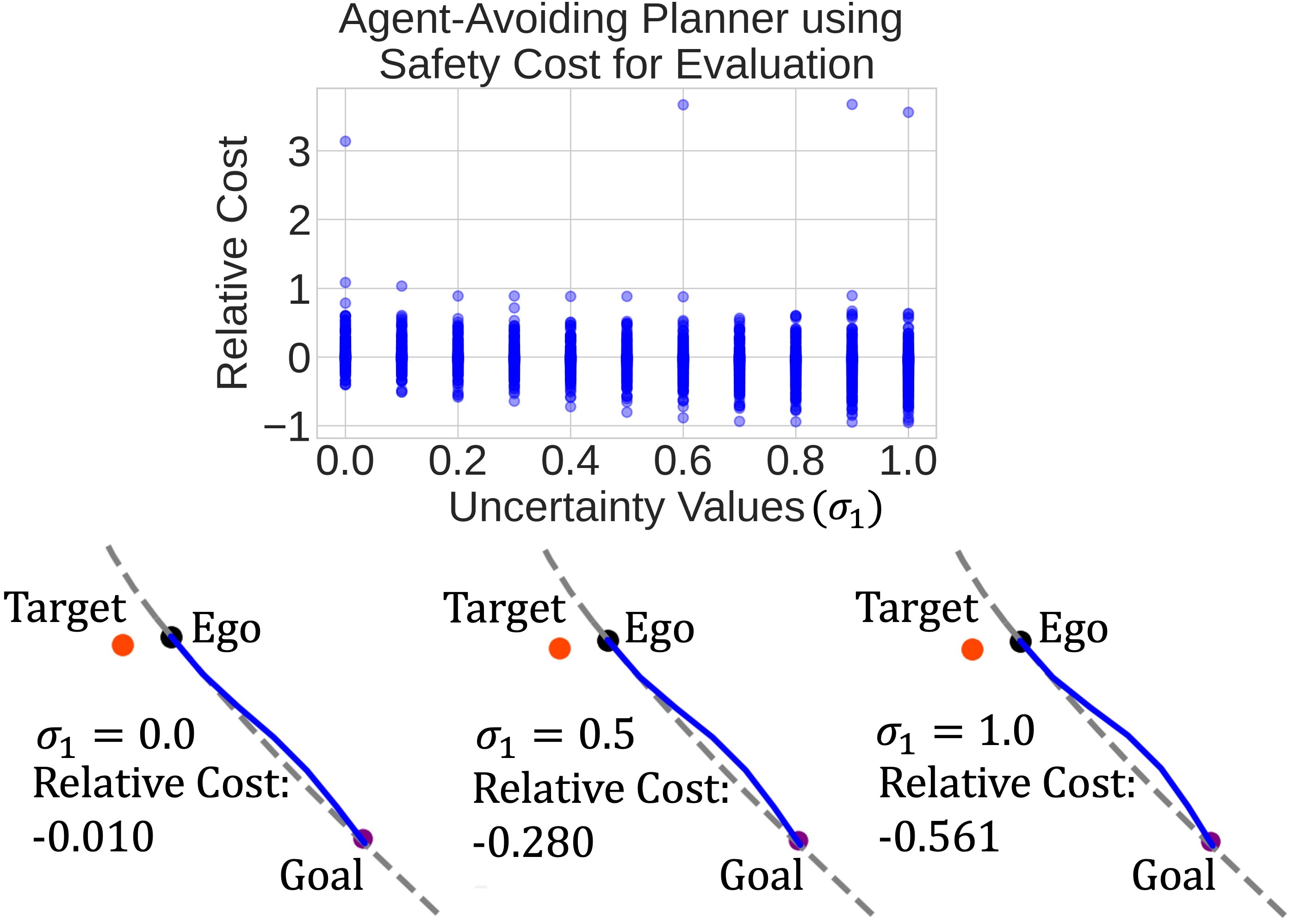}
\caption{(Running Example) (Top) A plot showing the raw data used to estimate $P(C(u) > c | \sigma_1, \mathbf{1}(e_1 < e_{thres}), z)$ for when the prediction error is below the threshold and the ego agent and target agent are close together.
(Bottom) Plots showing one of the scenarios in the dataset with the associated plan for three different settings of the uncertainty value.
Since this is the agent-avoiding planner, we indeed see that as the uncertainty value increases, the plan (dark blue solid line) takes a more conservative maneuver to reach the goal.
The evaluation cost here is the safety cost.}
\label{fig:raw_data_and_scenarios}
\end{figure}

\textbf{Criteria for Choosing the $G_1$ Object:} \\
(a) $\mathcal{G}_1$ should be related to the uncertainty measure $\sigma_i$ in some way. \\
(b) It must allow for tractable and interpretable analysis. \\
(c) It must be such that $G_1 \supseteq A_2$ for all $G_1 \in \mathcal{G}_1$ and $A_2 \in \mathcal{A}_2$. 

Given these criteria, the designer must choose $\mathcal{G}_1$ depending on the problem setting.
In particular, Criterion (c) indicates that we must consider what factors the downstream modules from module $i$ depend on for their performance.
Those factors then help to dictate what $\mathcal{A}_2$ is.
In our analyses, we have found that these factors are often related to the type of environmental scenario $z$ and the relationship between the uncertainty measure $\sigma_i$ and module error $e_i$.

In our particular perspective, we define $A_2$ as being a set of distributions $P(g_{enc}(\sigma_i, e_i) | z)$, where $g_{enc}$ is a function that encodes $\sigma_i$ and $e_i$ in a way that captures their relationship.
Thus, $\mathcal{A}_2$ is the space of sets of distributions of the form given.
Now, we set $G_1$ equal to $A_2$ (and therefore $\mathcal{G}_1 = \mathcal{A}_2$).

\textit{\textbf{Running Example:} As mentioned above, the performance of the planner and controller depends on the type of environmental scenario and the relationship between the trajectory predictor's uncertainty value and its error.
Thus, we write $A_2$ as a set of distributions $P(\sigma_1, \mathbf{1}(e_1 < e_{thres}) | z)$, and $A_2 = G_1$. 
So, $g_{enc}(\sigma_1, e_1)$ here outputs $\{\sigma_1, \mathbf{1}(e_1 < e_{thres})\}$ for an error threshold $e_{thres}$.
$z$ is defined as $\mathbf{1}(distance(ego, target) < d_{thres})$ for a distance threshold $d_{thres}$. 
Note that $G_1$ is still unknown; we have simply defined what kind of object it is.
The specification that we will generate will delineate which distributions $P(\sigma_1, \mathbf{1}(e_1 < e_{thres}) | z)$ belong in $G_1$. 
}

We now write the specification by first using the Law of Total Probability.
We abbreviate $g_{enc}(\sigma_i, e_i)$ as $l_i$ and write

\begin{equation}\label{eqn:law_total_probability}
    P(C(u) > c) = \sum_z \sum_{l_i} P(C(u) > c | l_i, z)P(l_i | z)P(z) 
\end{equation}

We note that $l_i$ is a continuous random variable.
However, as we go forward we will rely on a discretization of $l_i$, thus justifying the use of the summation above.
The reasons for this are related to the ease of producing and using the specification, which will be explained further in Section \ref{sec:estimating_specification}.

Now, we use Equation (\ref{eqn:system_spec}) to incorporate the system-level specification:

\begin{equation}\label{eqn:final_spec}
    \sum_z \sum_{l_i} P(C(u) > c | l_i, z)P(l_i | z)P(z) < \alpha
\end{equation}

We see that if the $P(C(u) > c | l_i, z)$ and $P(z)$ values are known, this inequality produces a constraint on the values that $P(l_i | z)$, that is $P(g_{enc}(\sigma_i, e_i) | z)$, can take on.
In other words, Equation (\ref{eqn:final_spec}) is the desired specification over $P(l_i | z)$, giving the set of distributions of that form such that $P(C(u) > c) < \alpha$.
$G_1$ is therefore implicitly described by this inequality.

We connect back to the concept of the quotient contract by noting that the quantity $P(C(u) > c | l_i, z)$ describes how well the downstream decision-making-under-uncertainty modules perform given $\sigma_i$ and $e_i$.
That is, in Equation (\ref{eqn:final_spec}) we are using both knowledge of the system-level contract $(A_{sys}, G_{sys})$ and of how the downstream modules relate $A_2$ to $G_2$ to compute $G_1$\footnote{On a technical note, since $G_1$ is a set of distributions, it is not actually a property but a \textit{hyperproperty}, or a set of properties. However, the notion of the quotient can still apply by using the concept of \textit{hypercontracts} as introduced by Incer et al \cite{Incer:EECS-2022-99}, and the intuitions behind our formulation stay the same.}.

\textit{\textbf{Running Example:} Equation (\ref{eqn:final_spec}) provides a specification by determining values that the distribution $P(\sigma_1, \mathbf{1}(e_1 < e_{thres}) | z)$ can take; that is, by dictating how much probability mass can be placed for different values of $\sigma_1$ and $\mathbf{1}(e_1 < e_{thres})$ given different values of $z$.
$P(z)$ refers to the distribution that the designer can expect the environmental scenarios to follow with respect to the distance between the ego and target agent.
$P(C(u) > c | \sigma_1, \mathbf{1}(e_1 < e_{thres}), z)$ represents how the planner and controller react to the upstream trajectory predictor's uncertainty and error in terms of the final cost incurred. 
}

What remains is now to provide estimates of the $P(C(u) > c | l_i, z)$ and $P(z)$ values and point out how this aids in our analysis.
This will be discussed in the next two sections.
However, before we proceed we make the following remark.

\textbf{Remark 1 [Choosing $g_{enc}$ and $z$]:} We point out the importance and nuances behind the choice of $g_{enc}$ and $z$ that the designer makes for analysis.
Here, Criterion (b) plays a key role.
While the specification (\ref{eqn:final_spec}) is true no matter how $g_{enc}$ and $z$ are defined, the usefulness of the empirical analyses that we perform is dependent on these variables being interpretable and relevant to system performance.
For example, for interpretability we may desire that $C(u)$ not be independent of $g_{enc}(\sigma_i, e_i)$ and $z$.
However, if one simply sets $z$ equal to $x$ and $g_{enc}$ to the identity mapping, the specification (\ref{eqn:final_spec}) loses its meaning as constraining module properties.
Thus, overall our criteria provide the designer with much flexibility to define variables, though with these caveats.  

In our Case Study for the Runway Incursion Detection system in Section \ref{subsec:boeing_analysis_specifications}, we will show how $g_{enc}$ can be defined as a \textit{calibration error} function, and explain the implications.
Indeed, in that case study we even remove dependencies on $z$.
In this way, our formulation operates in as much generality as possible.
Future work can explore additional ways to define these variables; for example, one can imagine defining $z$ using learned embeddings of data, similar to \cite{realgen}.

\subsection{Estimating Constants in the Specification}\label{sec:estimating_specification}

As mentioned in the previous section, a key part of our analysis lies in the estimation problem of finding the $P(C(u) > c | l_i, z)$ and $P(z)$ values.
We treat $l_i$ as a discrete random variable in this estimation for several reasons.
Firstly, by doing so we avoid making assumptions on the form of these distributions, for example by parametrization, for performing the estimation itself.
Secondly, we simplify the nature of the specification for the designer.
In many real-world scenarios, a designer will check if their distribution $P(l_i | z)$ satisfies the specification by taking samples of the input $x$ to the system and running the desired modules.
In such cases, it is most natural to calculate the $P(l_i | z)$ values by performing counting per $l_i$ bin. 
Also, if $l_i$ is discrete the $P(C(u) > c | l_i, z)$ and $P(z)$ values have a nice interpretation of being weights to the $P(l_i | z)$ values in the specification.
This has implications for better understanding the efficacy of the decision-making-under-uncertainty modules that are using the uncertainty measure from module $i$, as we will show.

We first address the issue of the $P(z)$ values. 
These are estimates of the relative frequency of the types of scenarios that the system will encounter.
This can be produced by examining a representative dataset.
At design time, these estimates serve to weigh the $P(C(u) > c | l_i, z)$ values, indicating how important it is for the decision-making-under-uncertainty modules to perform well for different situations.

For the estimation of the $P(C(u) > c | l_i, z)$ values, we similarly use a dataset of inputs to the system and environmental scenarios.
We use this dataset to sample outputs from the decision-making-under-uncertainty modules, given input $y_i$ and $\sigma_i$ values.
The cost function can then be evaluated for every output.
We discretize $l_i$ by creating bins and rounding values into the bins. 
Then, for each bin value of $l_i$ and each value of $z$, we can compute the probability that $C(u) > c$ using counting.

\textit{\textbf{Running Example:} We estimate the $P(C(u) > c | \sigma_1, \mathbf{1}(e_1 < e_{thres}), z)$ values, representing the action of the planner and controller, using the following method, though other methods are also possible.
We first sort the dataset into scenarios in which the ego and target agent are close together and those in which they are far apart.
For the trajectory predictor error, we use the ADE (average displacement error) between the predicted most likely trajectory of the target agent and the ground truth trajectory.
For a threshold $e_{thres}$ on this value, we consider data points for which the trajectory predictor produces predictions with error less than and also greater than this value.
Then, we grid the space between $0$ and $1$ and select each grid point to be a $\sigma_1$ sample.
For each data point, we run the planner and controller for each $\sigma_1$ value and for the fixed trajectory prediction. 
A sample plot showing $\sigma_1$ plotted against $C(u)$ is shown in Figure \ref{fig:raw_data_and_scenarios} for the safety cost for one selected value of $\mathbf{1}(e_1 < e_{thres})$ and $z$.
Finally, we show plots in Figure \ref{fig:running_example_contracts} that give the final $P(C(u) > c | \sigma_1, \mathbf{1}(e_1 < e_{thres}), z)$ values for all $\mathbf{1}(e_1 < e_{thres})$ and $z$ value combinations.
Here, for our cost we use the relative cost; that is, the cost of the trajectory produced by the system minus that produced by a system with access to the ground truth target agent future.
}

One important observation from our approach is that our specification is data-driven and therefore approximate.
As the data that is used for the probability estimations becomes more representative of the system and environmental scenarios, the accuracy of the approximation also increases.
The burden is therefore on the designer to produce good estimates.
Another note that we make is on the discretization strategy that we have employed.
In practice, the $l_i$ bin sizes should be small to produce a fine discretization of the continuous-valued $l_i$ but also large enough to ensure that there are enough samples in each bin.
In future work, we would like to perform more rigorous statistical analysis that takes into account these approximations in order to make our specification more robust.
For instance, a distributionally robust specification can be developed, or a specification that accounts for the variance of each probability estimate.

\textbf{Remark 2 [Computational Complexity]:} Overall, the computational complexity of our technique comes from the estimation of the $P(C(u) > c | l_i, z)$ values, which requires a dataset, possibly a way to simulate the decision-making-under-uncertainty modules for different inputs, and an evaluation of the cost function for each data point.
Thus, the computation time is linear in the size of the dataset and the runtime of the decision-making forward pass and cost evaluation.
Given a fixed computational budget, this induces a tradeoff dictating how many data points can be processed.

\subsection{Discussion}\label{subsec:specifications_discussion}

\begin{figure*}[!t]
\centering
\subfloat[]{\includegraphics[width=3.5in]{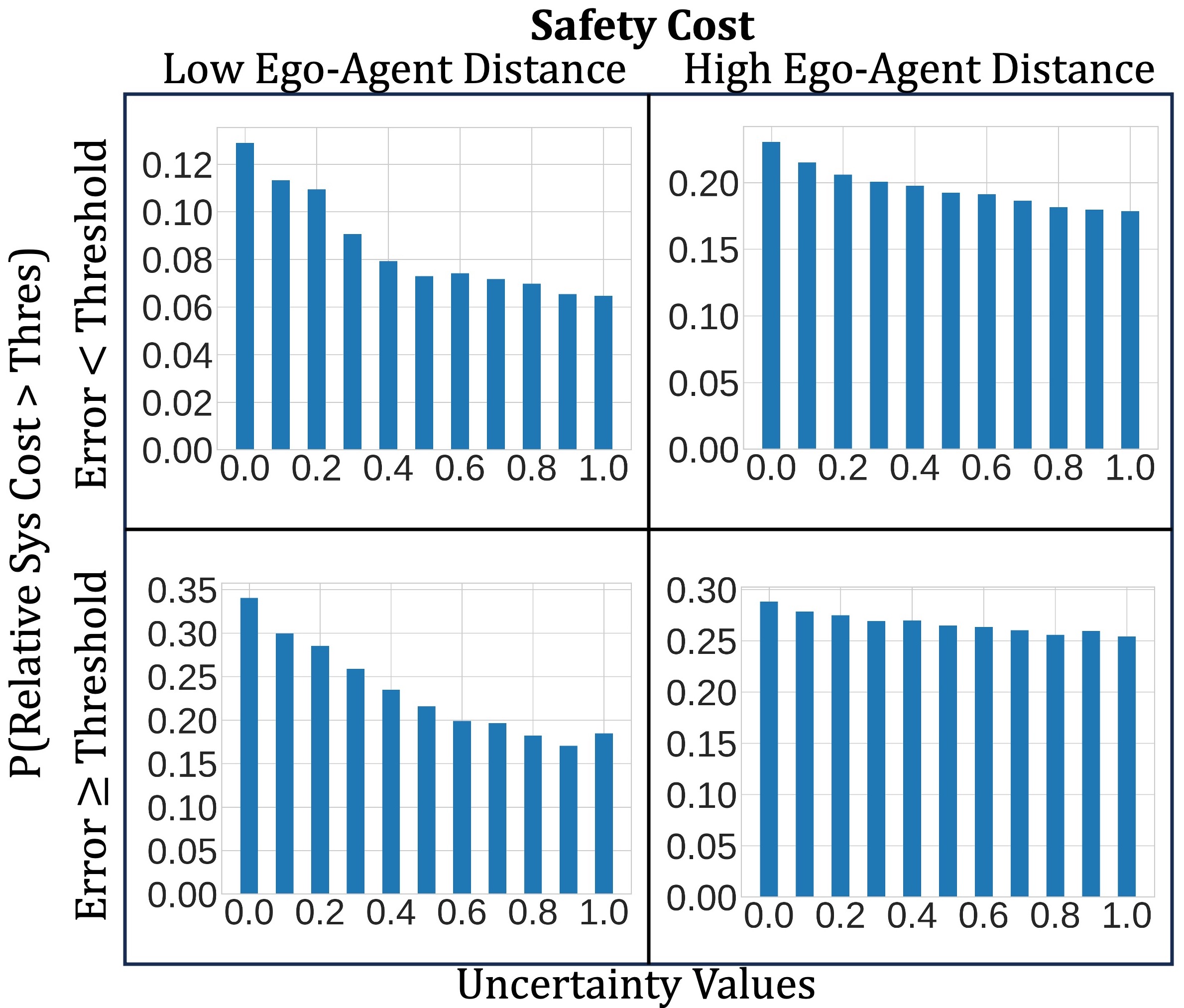}%
\label{fig_first_case}}
\hfil
\subfloat[]{\includegraphics[width=3.52in]{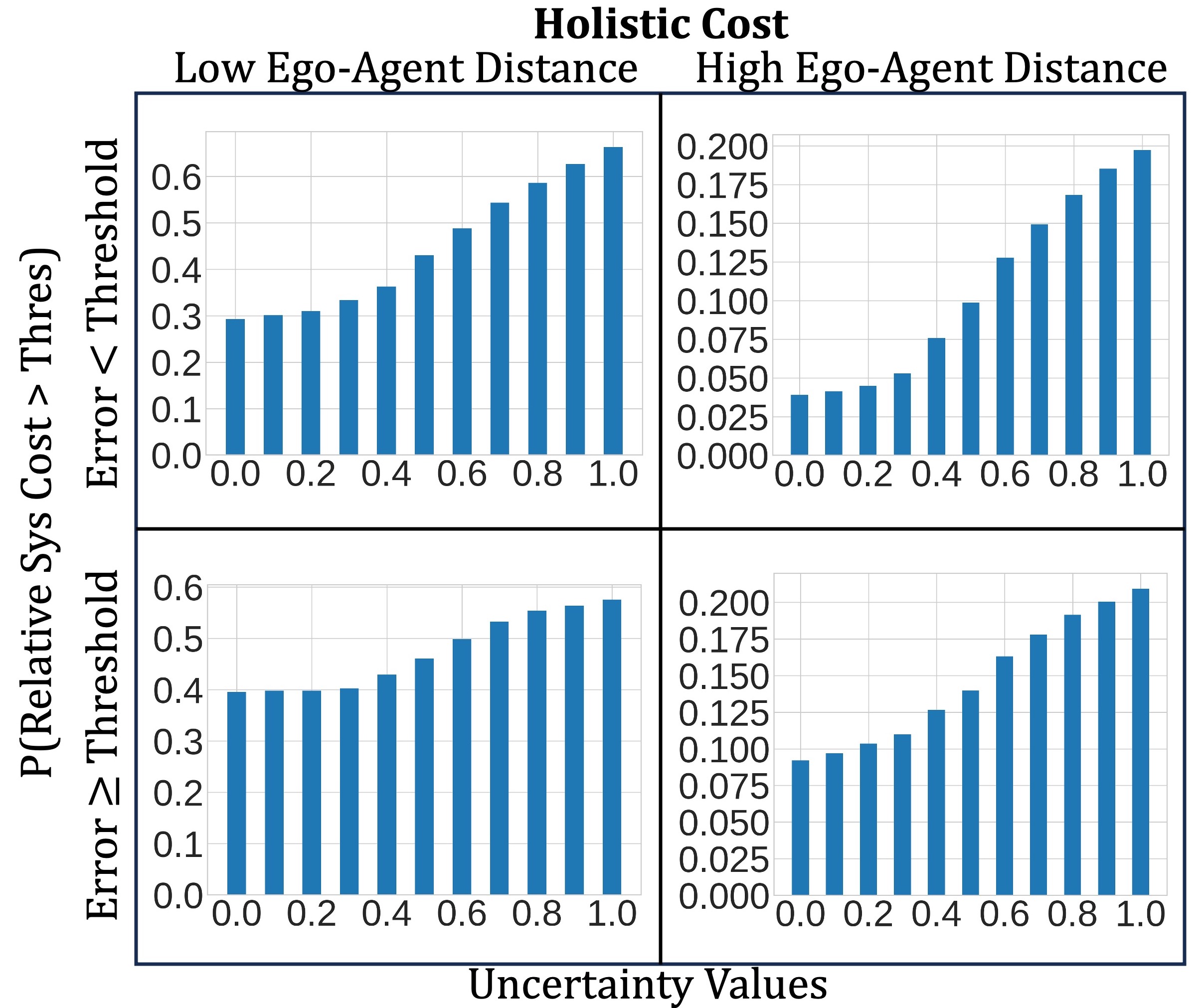}%
\label{fig_second_case}}
\caption{(Running Example) For the agent-avoiding planner, we show the final $P(C(u) > c | \sigma_1, \mathbf{1}(e_1 < e_{thres}), z)$ values for (a) evaluation using the safety cost and (b) evaluation using the holistic cost.
For each plot, we set the values of $\mathbf{1}(e_1 < e_{thres})$ and $z$ and show how the probability values vary by the uncertainty value.
The value of $e_{thres}$ is 1.5, the value of $c$ is 0.02, and a low ego-agent distance is defined as a distance less than 5 meters, while a high ego-agent distance is a distance greater than 5 meters.
In general, these values can be chosen arbitrarily by the system designer.
As the uncertainty value increases, the agent-avoiding planner keeps the ego agent further away from other agents, which can make the ego safer but worse at holistic driving.
}
\label{fig:running_example_contracts}
\end{figure*}

In this section, we discuss the implications that this specification generation process has on understanding uncertainty quantification and decision-making under uncertainty.
Firstly, the specification informs the system designer on the set of permissible distributions that govern a particular module's uncertainty measure such that the overall system performs well in probability.
In our formulation, this distribution relates the uncertainty measure with module error in some way.
So, the designer is informed of how much effort is needed to select or calibrate an uncertainty measure depending on how restrictive the specification is.

Secondly, the $P(C(u) > c | l_i, z)$ weights in the specification can be examined to yield additional insights on \textit{how well} the uncertainty measure is used by the downstream modules.
For example, if $l_i$ encodes a kind of distance between the uncertainty measure and the module error, thus capturing how well the uncertainty measure correlates with error, then the $P(C(u) > c | l_i, z)$ weights show the downstream modules' sensitivity to that value.
If the $P(C(u) > c | l_i, z)$ weight is high when $l_i$ is high, this indicates that the downstream modules are very sensitive to the correlation between the uncertainty measure and the error, liable to yield worse performance.
However, if the $P(C(u) > c | l_i, z)$ weight is low even when $l_i$ is high, this may indicate a greater robustness of the downstream modules to the upstream uncertainty measure's incorrect ability to predict the error.
Roughly speaking, in such a case the downstream modules can be thought to be ``calibrating'' the uncertainty measure themselves.
We illuminate these insights further with our running example.

\textit{\textbf{Running Example:} Again, the type of uncertainty-aware planner that we use is the agent-avoiding planner, whose internal cost function incentivizes candidate plans that stay away from other agents by an amount proportional to the uncertainty value.
We first reason about what our analysis should give us at an intuitive level, and then confirm the intuition using the formulation we have proposed.
}

\textit{First, we consider using the safety cost for evaluating the system.
In response to higher uncertainty values, the agent-avoiding planner prioritizes keeping a distance from the target agent and so will yield lower-cost plans in general.
If the prediction has a high error, the agent-avoiding planner will have the wrong reference trajectory to keep away from.
Thus, for higher error values we can expect higher-cost plans.
}

\textit{For the holistic cost, we can expect some of the opposite trends.
As the agent-avoiding planner prioritizes staying away from the target agent, it may de-prioritize other behaviors important for driving, such as lane-keeping and reaching the goal.
Thus, as the uncertainty value increases the planner will yield higher cost plans.
Similar to the situation for the safety cost, as the prediction error increases we can also expect the costs of the plans to increase.
}

\textit{Examining Figure \ref{fig:running_example_contracts}, we can indeed see the trends that we predict.
The heights of the bars in the plots that show $P(C(u) > c | \sigma_1, \mathbf{1}(e_1 < e_{thres}), z)$ values indicate how well the planner performs under different conditions and therefore show its sensitivity or ``robustness''.
For example, we notice that for the holistic cost evaluation, these bar heights decrease as the distance between the ego and target agent increases.
This is because when the distance is high, there is less need for the agent-avoiding planner to perform as aggressive avoiding maneuvers and so its holistic driving cost does not increase as much.
Thus, there is more sensitivity of the planner to scenarios with smaller ego-target agent distances.
It is also interesting to note that for the holistic evaluation cost, when the prediction errors are high the bar values tend to be higher for the lower uncertainty values compared to when the prediction errors are low.
There seems to be an interpretation here that the planner is also sensitive to how well the uncertainty value corresponds with the error -- if higher error values do not correspond with higher uncertainty values as well, the planner does not perform as well.
}

\textit{For the safety evaluation cost, we see that the bar heights actually increase as the ego-agent distance increases.
Often, for scenarios with high ego-agent distances, the planner does not prioritize avoidance as much and so outputs plans that are closer to the other agent and therefore less safe.
We also notice that when the error is higher, the bar heights decrease as the ego-agent distance increases. 
This is indicative of the greater sensitivity of the planner to higher errors, but only when the ego and target agent are in close range.
Otherwise, we still see the trend of higher bar values for higher errors when $z$ is fixed.
}

\textit{A final remark we make here is in noting the opposing trends between using the safety cost and the holistic cost for evaluation.
We consider the $P(\sigma_1, \mathbf{1}(e_1 < e_{thres}) | z)$ distributions that satisfy the final specifications.
For $\sigma_1$ values for which the $P(C(u) > c | \sigma_1, \mathbf{1}(e_1 < e_{thres}), z)$ weight is high, the $P(\sigma_1, \mathbf{1}(e_1 < e_{thres}) | z)$ distribution must have less probability mass in order for it to satisfy the specification.
Similarly, the $P(\sigma_1, \mathbf{1}(e_1 < e_{thres}) | z)$ distribution can have more probability mass when the $P(C(u) > c | \sigma_1, \mathbf{1}(e_1 < e_{thres}), z)$ weight is low.
This means that for the safety cost, higher uncertainty values are permissible with higher frequency, whereas for the holistic cost, the opposite is true.
We take this as an indication of the tradeoffs that can be present in decision-making under uncertainty.}

\subsection{Results on Case Studies} \label{subsec:specifications_results_case_studies}

\begin{figure*}[!t]
\centering
\subfloat[]{\includegraphics[width=3.52in]{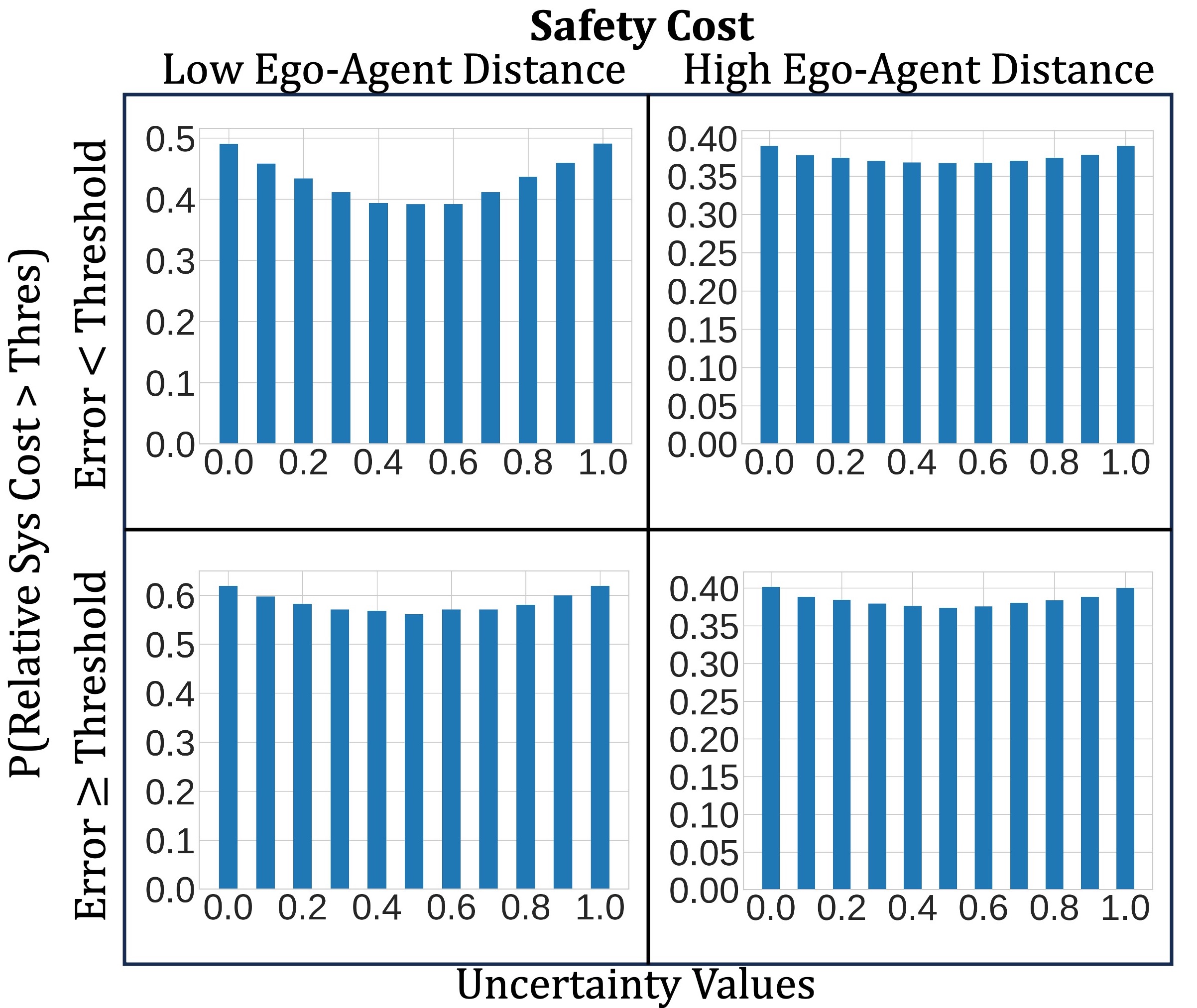}%
\label{fig_first_case}}
\hfil
\subfloat[]{\includegraphics[width=3.5in]{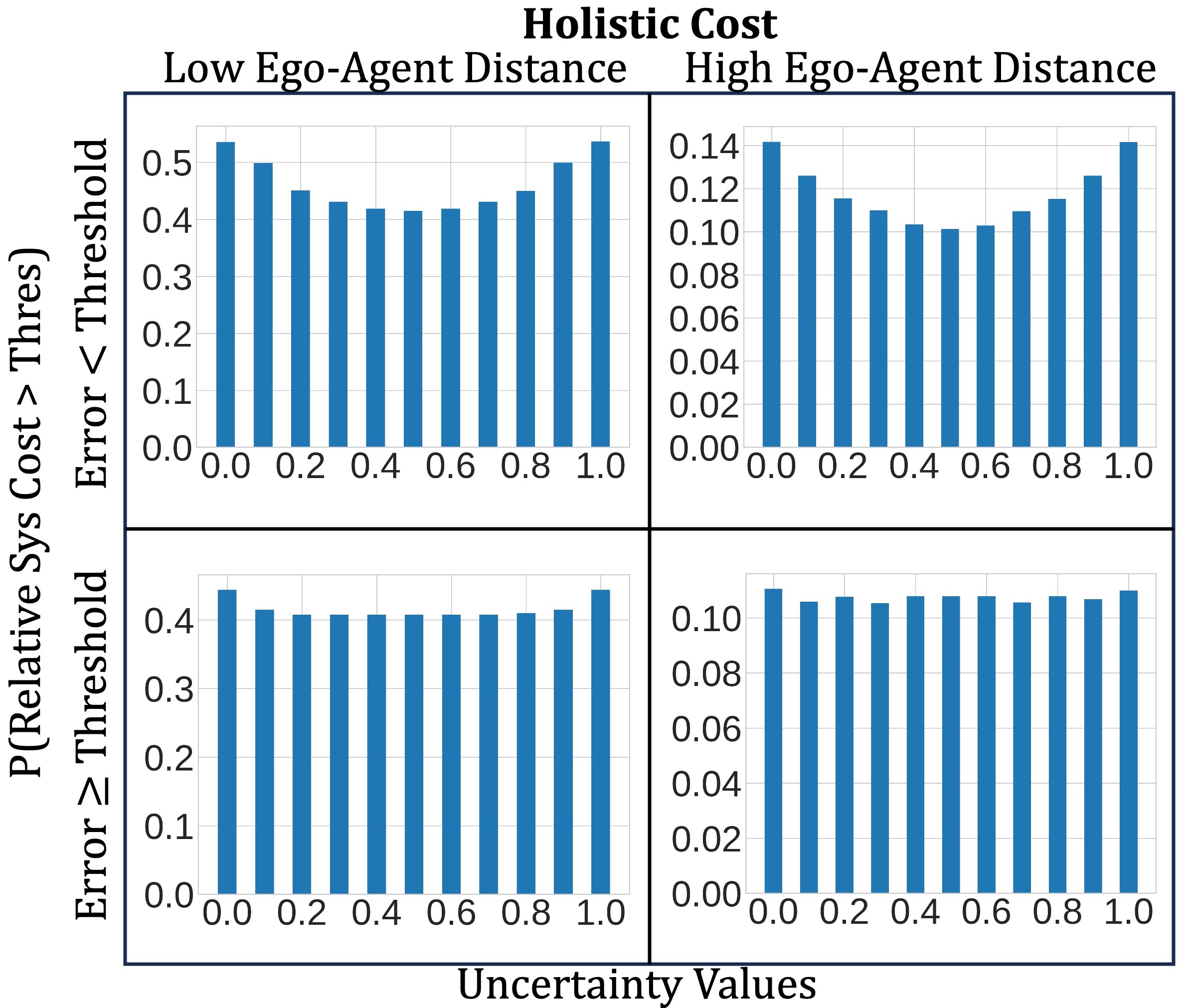}%
\label{fig_second_case}}
\caption{(Autonomous Driving System Case Study) For the lane-keeping planner, we show the final $P(C(u) > c | \sigma_1, \mathbf{1}(e_1 < e_{thres}), z)$ values for (a) evaluation using the safety cost and (b) evaluation using the holistic cost.
For each plot, we set the values of $\mathbf{1}(e_1 < e_{thres})$ and $z$ and show how the probability values vary by the uncertainty value.
Again, the value of $e_{thres}$ is 1.5, the value of $c$ is 0.02, and a low ego-agent distance is defined as a distance less than 5 meters, while a high ego-agent distance is a distance greater than 5 meters.
As the uncertainty value increases, the lane-keeping planner's cost does not monotonically increase for either the safety or holistic costs, since the planner is using the uncertainty value to trade off between different objectives.}
\label{fig:lanekeep_planner_chart}
\end{figure*}

\subsubsection{Autonomous Driving System}\label{subsec:nvidia_analysis_specifications}

In this section, we show the results of our analysis for the Autonomous Driving system. 
Since we have already performed our analysis for the agent-avoiding planner as part of the Running Example in the previous section, here we show the analysis for the lane-keeping planner (using the internal cost function given by Eq. (\ref{eq:cint_lane})).
We keep the same choice of $g_{enc}$ and $z$ and the same technique for estimating the $P(C(u) > c | \sigma_1, \mathbf{1}(e_1 < e_{thres}), z)$ values as for the agent-avoiding planner.
For both of these analyses, there are 12,452 data points in our dataset and it takes on the order of 15 minutes to obtain results on this dataset per $\sigma_1$ grid point using a GPU.
Figure \ref{fig:lanekeep_planner_chart} shows our results.

One of the first things to notice is the interesting U-shaped trend that we see in the $P(C(u) > c | \sigma_1, \mathbf{1}(e_1 < e_{thres}), z)$ values as $\sigma_1$ increases for most of the plots, though this trend is not as present for when the predictor error is high and the ego and target agent are far from each other.
Although the shape is not perfectly symmetric, it suggests that there is an intermediate value of $\sigma_1$, roughly $0.5$, at which the planner operates best for both the safety and holistic costs.
If the uncertainty value is too low, the ego agent may not prioritize goal-reaching and collision avoidance enough, which results in a higher cost in terms of both safety and holistic driving.
If the uncertainty value is too high, the ego agent may excessively prioritize lane-keeping, causing it to not maintain a safe enough distance from other agents and also de-prioritize goal-reaching.
Thus, it seems important for the planner to be able to effectively balance all five of the internal cost objectives given in Equation (\ref{eq:internal_c}).
In this way, the particular function we choose for $C_{\text{lane}}^{\text{int}}$ matters greatly; for example, when $\sigma_1 = 0$ the planner actually prioritizes lane-keeping more than when $\sigma_1 = 0.1$. 
When $\sigma_1 = 0.5$, for some data points the planner performs the maximally agent-avoiding maneuver.
Clearly, a different method of weighting the planner internal cost terms would have demonstrated different behaviors.

Similar to the Running Example, we see trends in which the bar heights increase for the safety evaluation cost as errors increase when $z$ is fixed.
For both the safety and holistic evaluation costs, the bar heights decrease when the ego-agent distance increases.
Since the bar heights for this planner are high in the first place compared to those for the agent-avoiding planner, this may be simply due to the fact that high ego-agent distances decrease cost values in general.
For the holistic evaluation cost, the bar heights actually decrease as errors increase when $z$ is fixed.
This is an interesting phenomenon, which we hypothesize points to the effectiveness of a general lane-keeping strategy in the presence of higher prediction errors.
Indeed, we see this effect in our robustness analysis as well, in Section \ref{subsubsec:robustness_av_sys}.

\subsubsection{Runway Incursion Detection System} \label{subsec:boeing_analysis_specifications}

We now pivot to discussing the Runway Incursion Detection system described in Section \ref{sec:boeing_system_description}.
We focus on $\sigma_1$, which represents the sequence of detector confidence scores corresponding to all of the detections made of a tracked object during the track's existence.
Note that this means that $\sigma_1$ does not have a fixed dimension and can include fairly long sequences of numbers, for example for a track that lasts a long time.

The key elements for the setup of this case study lie in the choice of $g_{enc}$ and $z$.
Although this system may operate in different categories of environmental scenarios, we follow Remark 1 and Criterion (c) and find that the tracker performance can be explained solely using a particular choice of $g_{enc}$ and without the help of any $z$. 
For $g_{enc}$, we follow Criterion (b) and set it to represent the \textit{calibration error} of the detector for a particular track.
Calibration error can be computed in multiple ways, for example by considering the Expected Calibration Error (ECE) \cite{naeini_ece}. 
Here, since we work with a bounding-box detector neural network, we compute the Detection Expected Calibration Error (D-ECE) \cite{Kppers2020MultivariateCC} per track: 

\begin{equation}\label{eqn:dece}
    \sum_{j=1}^{J_{total}} \frac{|I(j)|}{N} \times |\text{avg precision}(j) - \text{avg conf}(j)|
\end{equation}
where we follow \cite{Kppers2020MultivariateCC} in partitioning the detector confidence and bounding box property spaces into bins.
The total number of bins is $J_{total}$ and $j$ represents the index of each bin.
Here, $I(j)$ is the set of detections that are in a single bin.
$N$ is the total number of detections in the track.
The precision is a binary value representing whether the detection is correct or not, where correctness is determined by matching the detection with a ground-truth bounding box with sufficient IoU (Intersection over Union).
The averages are taken per bin.
In our analysis, we perform the binning by rounding values.
Similar to the considerations for binning the $l_i$ space as described in Section \ref{sec:estimating_specification}, the choices of bin sizes here are also determined by a tradeoff between the accuracy of the computation and how many samples lie in each bin. 
Here, we set the bin size for the confidences to 0.05 and the bin size of the box property space to 50.

\begin{figure}[!t]
\centering
\includegraphics[width=3.5in]{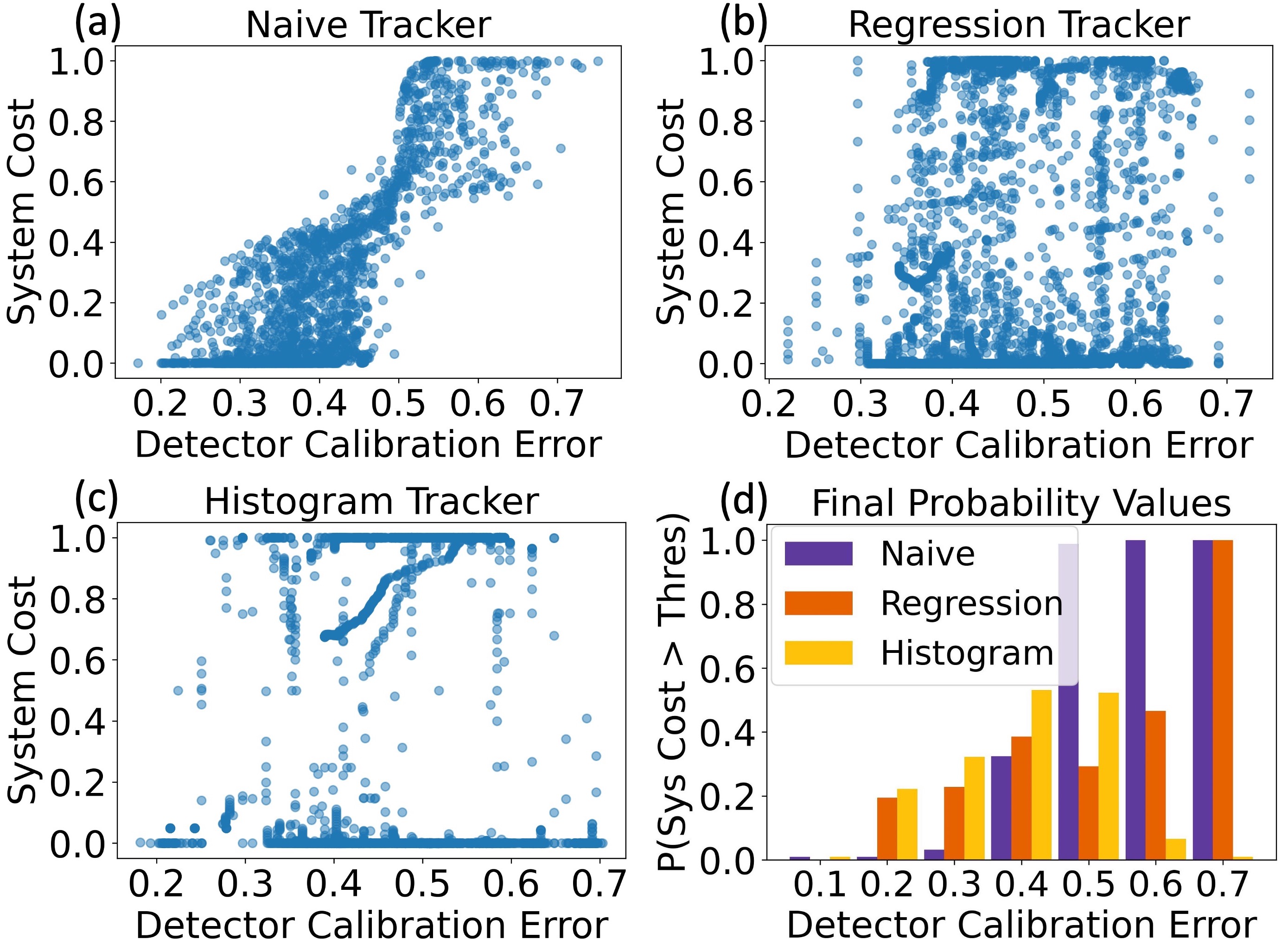}
\caption{\label{fig:boeing_raw_data} (Runway Incursion Detection System Case Study) We show the data used to estimate the $P(C(u) > c | l_1)$ values for the Naive Tracker (a), Regression Tracker (b), and Histogram Tracker (c).
Each point represents a track, which corresponds to a sequence of detector confidence values.
These points are binned along the Calibration Error ($l_1$) axis.
(d) Plot showing the $P(C(u) > c | l_1)$ values for the Naive, Regression, and Histogram trackers side-by-side.
The lowest bars represent a value of $0$.
The value of $c$ is 0.4 and the detector calibration errors have bin sizes of 0.1.}
\end{figure}

For this Case Study, our specification is 

\begin{equation}\label{eqn:boeing_final_spec}
    \sum_{l_1} P(C(u) > c | l_1)P(l_1) < \alpha
\end{equation}

This specifies the values that $P(l_1)$ can take, thus informing the designer on how much effort needs to be expended in calibrating the detector's uncertainty measure.

The second part of the setup for this Case Study lies in describing how we estimate the $P(C(u) > c | l_1)$ values.
These values characterize the tracker module and how well it uses the uncertainty measure from the detector.
While for the Autonomous Driving system we could grid the uncertainty measure space directly to generate the samples, in this case the high and variable dimensionality of $\sigma_1$ makes this difficult.
Since information is lost by producing the encoding given by $g_{enc}$, which outputs a scalar, one cannot use samples of $l_1$ to produce $\sigma_1$ values and run the tracker.
Therefore, our estimation technique is to run the detector and tracker module together for each tracker design for several test conditions.
Here, a test condition is a particular aircraft landing scenario that lasts several minutes and that involves certain ground vehicles and results in a set of tracks being produced.
Combining the data from our set of 12 test conditions, we obtain a dataset of tracks with their associated confidence scores, detections, and ground-truth bounding boxes.
We compute $l_1$ for each track and plot these values against $C(u)$.
Then, we bin the points along the $l_1$ axis by rounding the $l_1$ values.
Finally, in each bin we can compute $P(C(u) > c | l_1)$ via counting.
In terms of computational complexity, although each test condition is short, it can be time intensive to compute these probabilities due to overhead in running the system in ROS.

In Figure \ref{fig:boeing_raw_data}, we show both the raw data plots that we use to estimate the $P(C(u) > c | l_1)$ values as well as the $P(C(u) > c | l_1)$ values themselves.
We perform these analyses for the Naive, Regression, and Histogram trackers.
Similar to our analyses for the Autonomous Driving system and Running Example, we can understand these plots in terms of what they indicate about the efficacy of the trackers and in terms of the specifications they generate.
We see that as $l_1$ increases, the $P(C(u) > c | l_1)$ values increase as well, most sharply for the Naive tracker. 
Thus, we can say that the Naive tracker, while it is the simplest computationally, is the most sensitive, or least robust, to miscalibration in the detector.
In comparison, the Regression and Histogram trackers produce lower $P(C(u) > c | l_1)$ weights.
These trackers are taking into account a sliding window of confidence values over the track and using more sophisticated functions to compute the track's probability of existence.
In this way, detector miscalibration does not affect these tracker designs as adversely.

As mentioned in our Running Example analysis, what this means for the generated specifications is that if the Naive tracker is being used downstream, less probability mass can be placed on higher values of $l_1$ for the $P(l_1)$ distribution.
However, if the Regression or Histogram trackers are being used instead, more probability mass is permissible for higher values of $l_1$ for the system specification to be satisfied.
In essence, using a more intelligent decision-making-under-uncertainty algorithm compensates for a more untrustworthy uncertainty measure.
This analysis therefore contributes to our argument that only evaluating the calibration level of a standalone module is insufficient; by appropriately designing the entire system, concerns about the reliability of an individual module can become less relevant.

\section{Measuring System Robustness} \label{sec:measuring_robustness}

\subsection{Method}

\begin{figure*}[!t]
    \vspace{0.0625in}
    \centering
    \subfloat[]{%
        \includegraphics[width=0.325\textwidth]{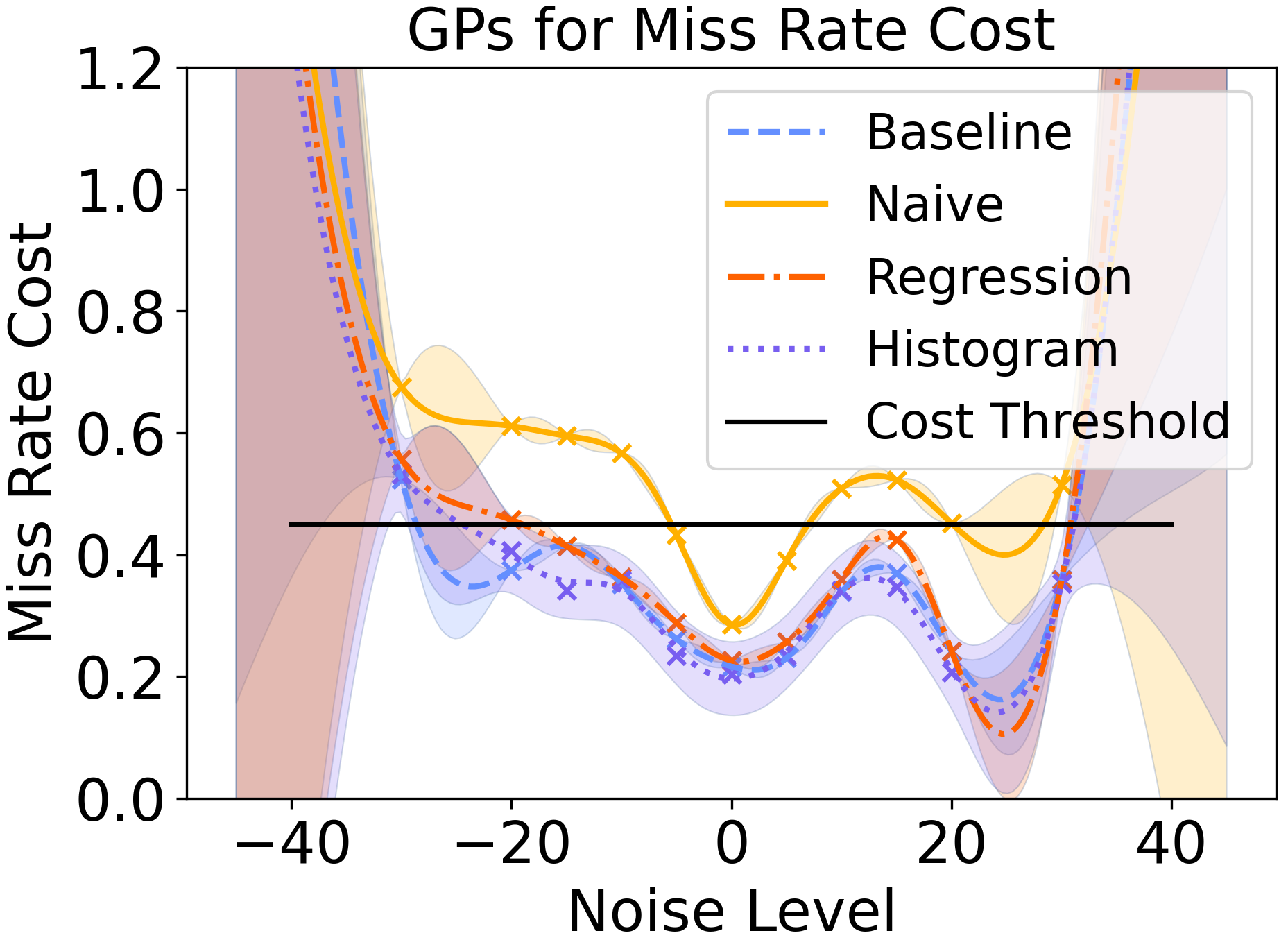}
    }
    \hfill
    \subfloat[]{%
        \includegraphics[width=0.32\textwidth]{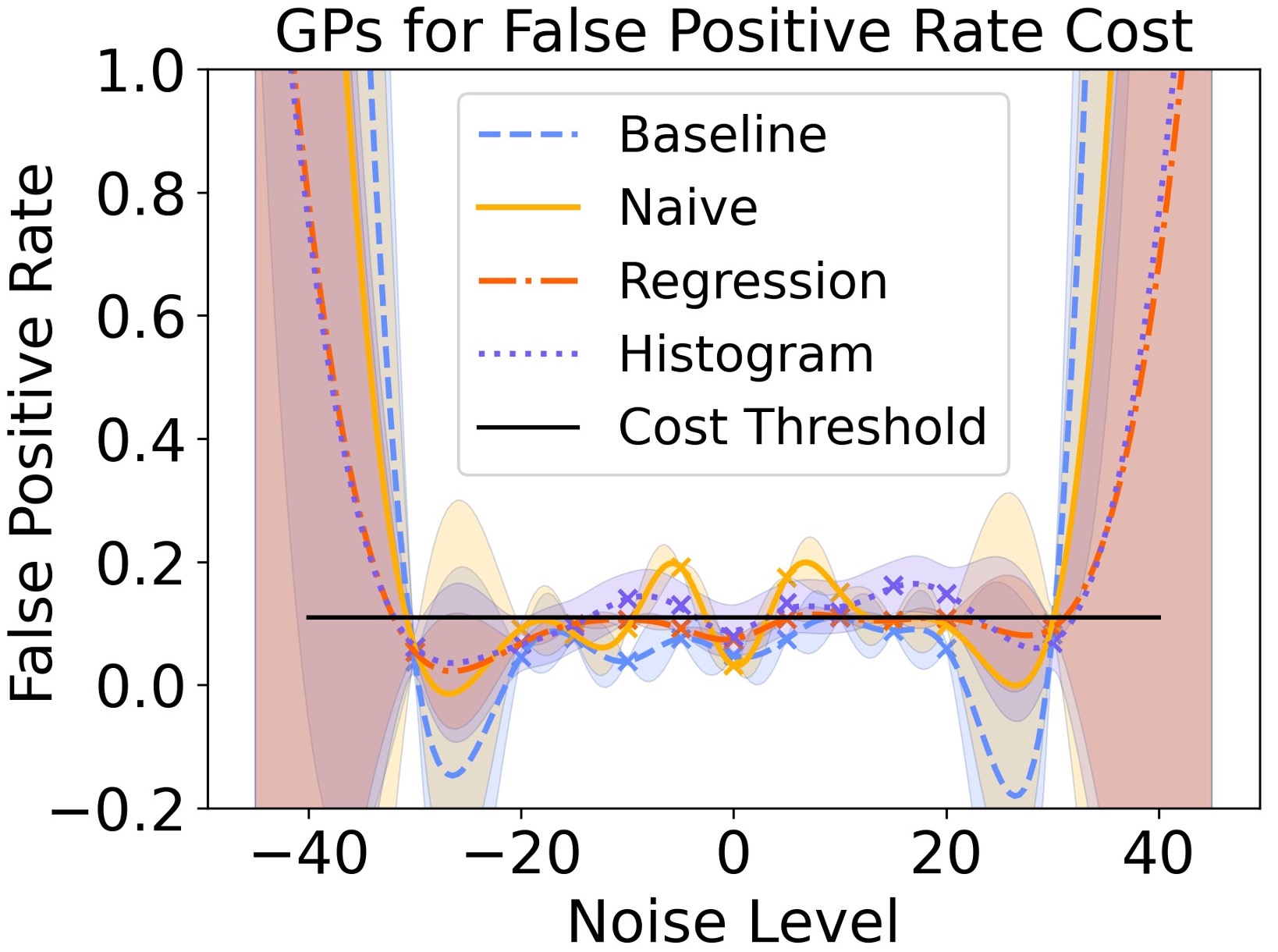}
    }
    \hfill
    \subfloat[]{%
        \includegraphics[width=0.325\textwidth]{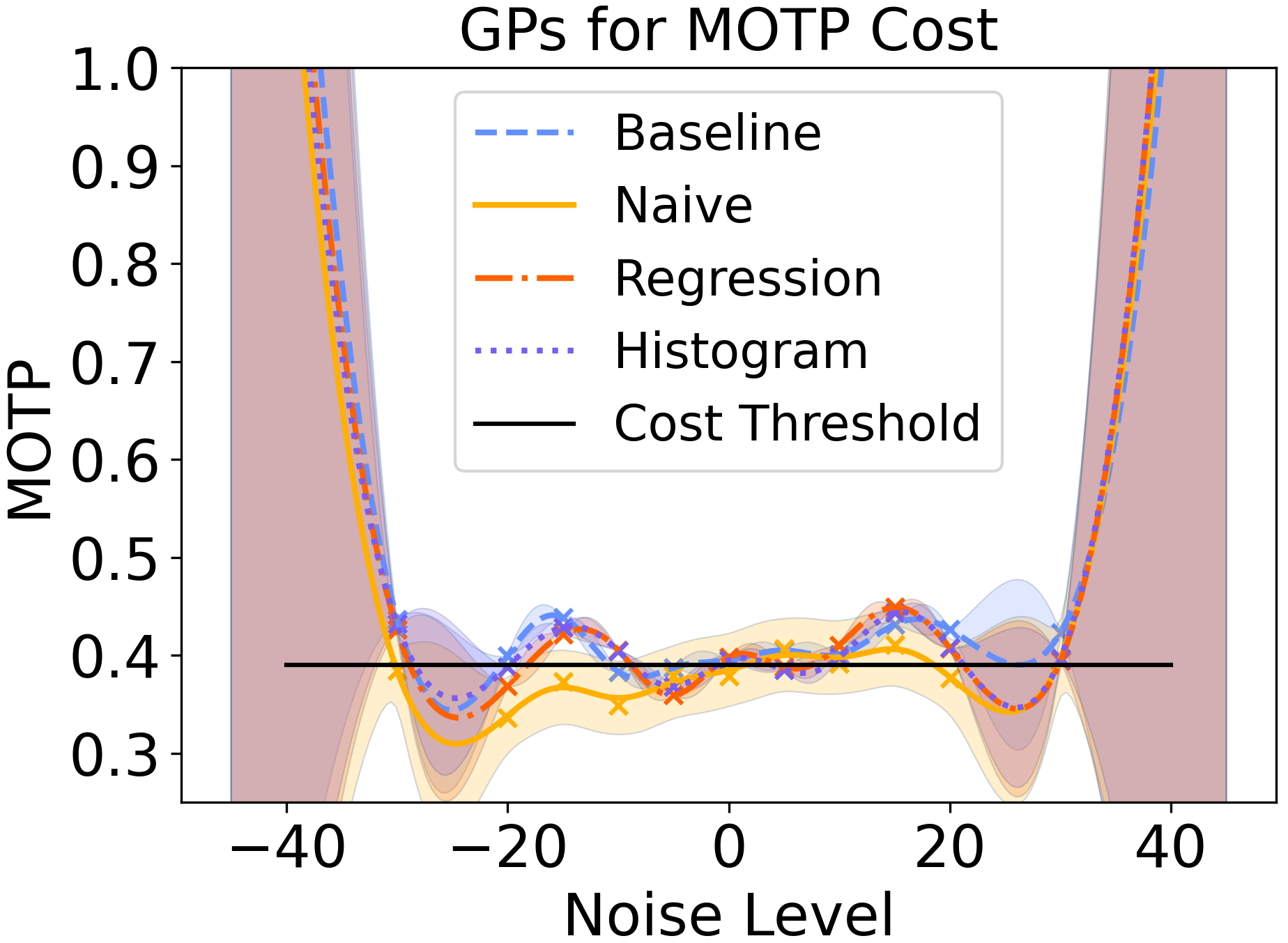}
    }
    \caption{\label{fig:boeing_gps} (Runway Incursion Detection System Case Study) We plot the fitted Gaussian Processes representing $F(\epsilon)$ for three different cost functions: the miss rate (a), false positive rate (b), and MOTP (c).
    Thus, here ``noise level'' refers to the $\epsilon$ brightness perturbation value.
    Computing the sub-level sets of these functions yields a one-dimensional interval, which we then compute the size of by counting grid points.}
    \vspace{-0.15in}
\end{figure*}

As described in Section \ref{sec:problem_robustness}, for our second analysis we wish to find the size of the set $\mathcal{S}$ as given in Equation (\ref{eqn:robustness_set}), repeated here for clarity.
This is the set of input perturbations under which the system is still performant on average:

\begin{equation*} 
    \mathcal{S} := \{\epsilon \mid \mathbb{E}_{x \sim \mathcal{D}} [(C \circ f)(p_{\epsilon}(x))] < c \}
\end{equation*}

In order to solve this problem, we observe that it is essentially a sub-level set estimation problem for the function $\mathbb{E}_{x \sim \mathcal{D}} [(C \circ f)(p_{\epsilon}(x))]$ with argument $\epsilon$.
Since we assume that $f$ is a black box, we perform this estimation via sampling.
That is, we sample $\epsilon$ values, and for each sample, we use samples of $x$ to estimate the expected value.
Similar to prior work \cite{Gotovos2013, Iwazaki2020, Inatsu2021} based on the principles of Bayesian optimization \cite{Frazier2018ATO}, we fit a Gaussian Process (GP) \cite{Rasmussen2006Gaussian} to the samples.
This gives us an estimate of the function $\mathbb{E}_{x \sim \mathcal{D}} [(C \circ f)(p_{\epsilon}(x))]$.
The fact that a GP models the stochasticity inherent in producing each sample allows for a probabilistic approach.
In particular, we can take advantage of an acquisition function to determine where to sample in the space of $\epsilon$ and thus potentially reduce the computational complexity of the problem.
In this work, we rely on the Maximum Improvement in Level-Set Estimation (MILE) acquisition function, introduced by Zanette et al \cite{Zanette2018RobustSS}.
Overall, this is one technique for performing sub-level set estimation, and other techniques may also be considered, especially if the statistic in the calculation is not the expected value, as was noted in Section \ref{sec:problem_robustness}.

Let us define $F(\epsilon)$ as $\mathbb{E}_{x \sim \mathcal{D}} [(C \circ f)(p_{\epsilon}(x))]$.
Then, we can define $\hat{F}(\epsilon) = \frac{1}{N}\sum_{j=1}^{N} (C \circ f)(p_{\epsilon}(x_j))$ as the sample mean for $N$ samples. 
Following one of the methodologies described by Katz et al \cite{katz2023}, we aim to find all $\epsilon$ such that

\begin{equation}\label{eqn:gp_probabilistic_guarantee}
    P(F(\epsilon) < c) > \gamma
\end{equation}
We model $F(\epsilon)$ as a Gaussian distributed random variable with mean $\hat{F}(\epsilon)$.
We then fit the Gaussian Process using samples $\{ \hat{F}(\epsilon_1),...\hat{F}(\epsilon_M)\}$ for $M$ samples.
Let $\mu_{GP}(\epsilon)$ and $\sigma_{GP}(\epsilon)$ be the resulting posterior mean and standard deviation functions. 
Then, we include a new point $\epsilon'$ into the sub-level set if $ \mu_{GP}(\epsilon') + \beta \sigma_{GP}(\epsilon') < c$, where $\beta$ is set such that a $\gamma$ amount of probability mass is included for a Gaussian distribution.
We find the size of $\mathcal{S}$ by gridding the space of $\epsilon$ and then counting the number of grid points that satisfy the above condition. 
We divide this number by the total number of grid points.

\textbf{Remark 3 [Computational Complexity]:} The majority of the computation time of the technique we have described lies in producing estimates of $\mathbb{E}_{x \sim \mathcal{D}} [(C \circ f)(p_{\epsilon}(x))]$ for each $\epsilon$ sample.
Thus, producing all of the samples for which we fit the GP takes time linear in $M$, $N$, and the simulation time for running $f$ on each $x$.
Again, this induces a tradeoff dictating the magnitude of these quantities.
In future work, we would like to explore how system knowledge might aid in reducing this computational complexity.
Once the GP is fit, some additional time, linear in the amount of grid points in the space of $\epsilon$, is taken to compute the size of $\mathcal{S}$.

\subsection{Results on Case Studies}

For each of our Case Studies, we follow certain steps in our analysis: (1) We first define $p_\epsilon(x)$, (2) We enumerate the different system designs that we would like to compute the robustness metric for, and (3) We compute the sub-level sets for each system design and examine each design in terms of both robustness and performance. \\

\subsubsection{Runway Incursion Detection System}

Recall that for this system, $x$ represents images of the runway from the aircraft camera.
Since for this portion of our analysis we are interested in having a comprehensive understanding of the system's input-output behavior, we modify our definitions of $x$ and $u$ slightly.
We have $x$ represent all images in a test condition, and $u$ represent all tracks produced during a test condition.
Since in our formulation we allow for the perturbation function $p_\epsilon(x)$ to capture any kind of semantic shift to the input, we can solve our sub-level set estimation problem in a potentially low-dimensional space.
In this Case Study, we set $p_\epsilon(x)$ to represent a constant pixel-wise addition to each image in $x$ by amount $\epsilon$, capped at the values of $0$ and $255$ for the RGB images.
This essentially changes the \textit{brightness} of the image; positive $\epsilon$ values increase the image brightness while negative $\epsilon$ values make the image darker.

The system designs that we will compare will all include the same detector module, but with different tracker modules.
We will compare the Naive, Regression, and Histogram trackers, as well as the tracker that does not use the uncertainty measure from the detector at all.
We will also consider different cost functions for evaluating the system output compared to Section \ref{subsec:boeing_analysis_specifications}.
We use the Miss Rate, False Positive Rate, and Multiple Object Tracking Precision (MOTP) as our three cost functions, which come from the tracking literature \cite{clearmot_metrics}.
The miss rate and false positive rate measure how often a track does not note the existence of an object and maintains an object that does not exist, respectively, with respect to the number of ground truth objects.
The MOTP measures how accurately the tracker maintains object positions over time with respect to ground-truth positions.

Figure \ref{fig:boeing_gps} shows the Gaussian Processes that we fit to represent $F(\epsilon)$ for each cost function.
For each $\epsilon$ sample, we sample test conditions 15 times, run all four systems for those samples, and compute the average cost per system.
In terms of computation, again although each test condition lasts less than five minutes, it is expensive to compute $F(\epsilon)$ due to overhead in running the system in ROS.
We take 11 samples of $\epsilon$ in total.
Since $\epsilon$ is only 1-dimensional in this case, we do not find it necessary to use the MILE acquisition function to achieve appropriate coverage of the space.
The cost threshold $c$ is an analysis choice and is set to 0.45 for the miss rate cost, 0.11 for the false positive rate cost, and 0.39 for the MOTP cost.
We set $\gamma$ from Equation (\ref{eqn:gp_probabilistic_guarantee}) to 0.9.
The GP priors have means that are specific quadratic functions and covariances specified by Matern52 kernel functions.
To calculate the sizes of the level sets, we grid the $\epsilon$ axis in increments of 0.5.

In Table \ref{tab:boeing_robustness_results}, we display the sizes of the sub-level sets of $F(\epsilon)$ in terms of the percent of grid points included in the sets, as well as the performance of each system design as measured by the cost value itself when $\epsilon = 0$.
In this way, we address our original intent of understanding how uncertainty-awareness affects a system's placement in the system design optimization landscape.
Clearly, a design may produce a good performance but lack in robustness, and vice versa.

Examining our results, we see that an uncertainty-aware design using the Histogram tracker performs better than the baseline no-uncertainty system in terms of miss rate performance. 
However, it as well as the Regression tracker design perform worse in terms of the false positive rate performance.
This is expected since the uncertainty-aware trackers are designed to be more conservative, maintaining tracks for longer periods of time and not requiring very high confidence score-detections to initiate a track in the first place.
This conservatism produces a tradeoff in terms of miss rate and false positive rate.
We also see that the Naive tracker performs the best in terms of MOTP.
This is because it only uses confidence scores at each timestep to determine whether to initiate or terminate a track, instead of using a history of confidence scores.
As a result, it often produces very short tracks, terminating a track once a confidence score drops instead of maintaining it for long periods.
We believe that this allows it to be more precise in estimating an object's location, as the tracked object's location matches more closely with the detections. 

In terms of our robustness results, we see that interestingly the baseline design has the highest sub-level set size for both the miss rate and false positive rate cost functions.
However, the uncertainty-aware designs have a higher robustness for the MOTP cost function.
One reason for this may be that as we add perturbations to the input image, the detector may be producing worse detections, for example by dropping detections or detecting non-objects, in a way that impacts trackers similarly.
Indeed, the data points for the Baseline and Histogram tracker system GPs tend to be close, at least for the miss rate cost.
If the detector produces the wrong detections, a tracker, even an uncertainty-aware one, may not be able to do much to ``recover'' and have a good performance.
However, in general we do see the presence of tradeoffs that we predicted -- while the Baseline system has advantages in some ways, it does not have a good MOTP robustness or performance.
Similarly, the Regression tracker system performs reasonably well in terms of miss rate and false positive rate robustness but poorly in terms of MOTP performance.

\begin{table}[] 
\centering 
\caption{Sub-Level Set Sizes and Costs for Tracker Designs} 
\label{tab:boeing_robustness_results} 
\hspace*{-0.3cm}
\begin{tabular}{|c|*{6}{c|}}

\hline

System & \multicolumn{2}{c|}{Miss Rate} & \multicolumn{2}{c|}{FP Rate} & \multicolumn{2}{c|}{MOTP} \\

\cline{2-7}

& Size of & Cost $\downarrow$ & Size of & Cost $\downarrow$ & Size of & Cost $\downarrow$ \\

& SL& \hfill & SL& \hfill & SL& \hfill  \\

& Set $\uparrow$& \hfill & Set $\uparrow$& \hfill & Set $\uparrow$& \hfill \\

\hline

Baseline & \textbf{0.927} & 0.217 & \textbf{0.798} & 0.048 & 0.0
 & 0.395 \\

\hline

Naive & 0.218 & 0.286 & 0.0 & \textbf{0.032} & \textbf{0.315}
 & \textbf{0.379} \\

\hline

Regression & 0.774 & 0.227 & 0.556 & 0.075 & 0.218
 & 0.398 \\

\hline

Histogram & 0.823 & \textbf{0.204} & 0.169
 & 0.079 & 0.185
 & 0.395 \\

\hline

\end{tabular}
\end{table}

\begin{figure}[!t]
\centering
\includegraphics[width=2.4in]{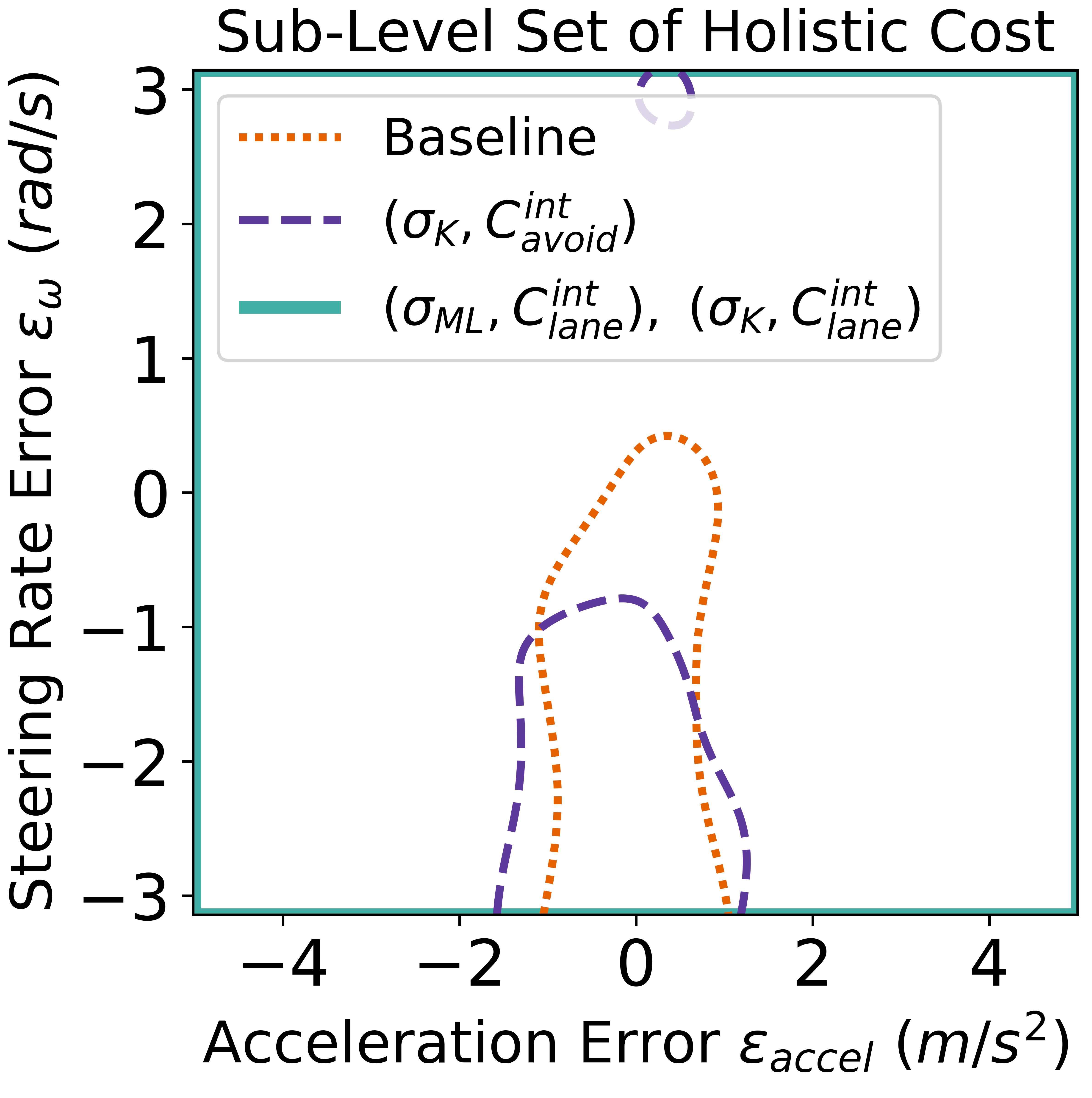}
\includegraphics[width=2.4in]{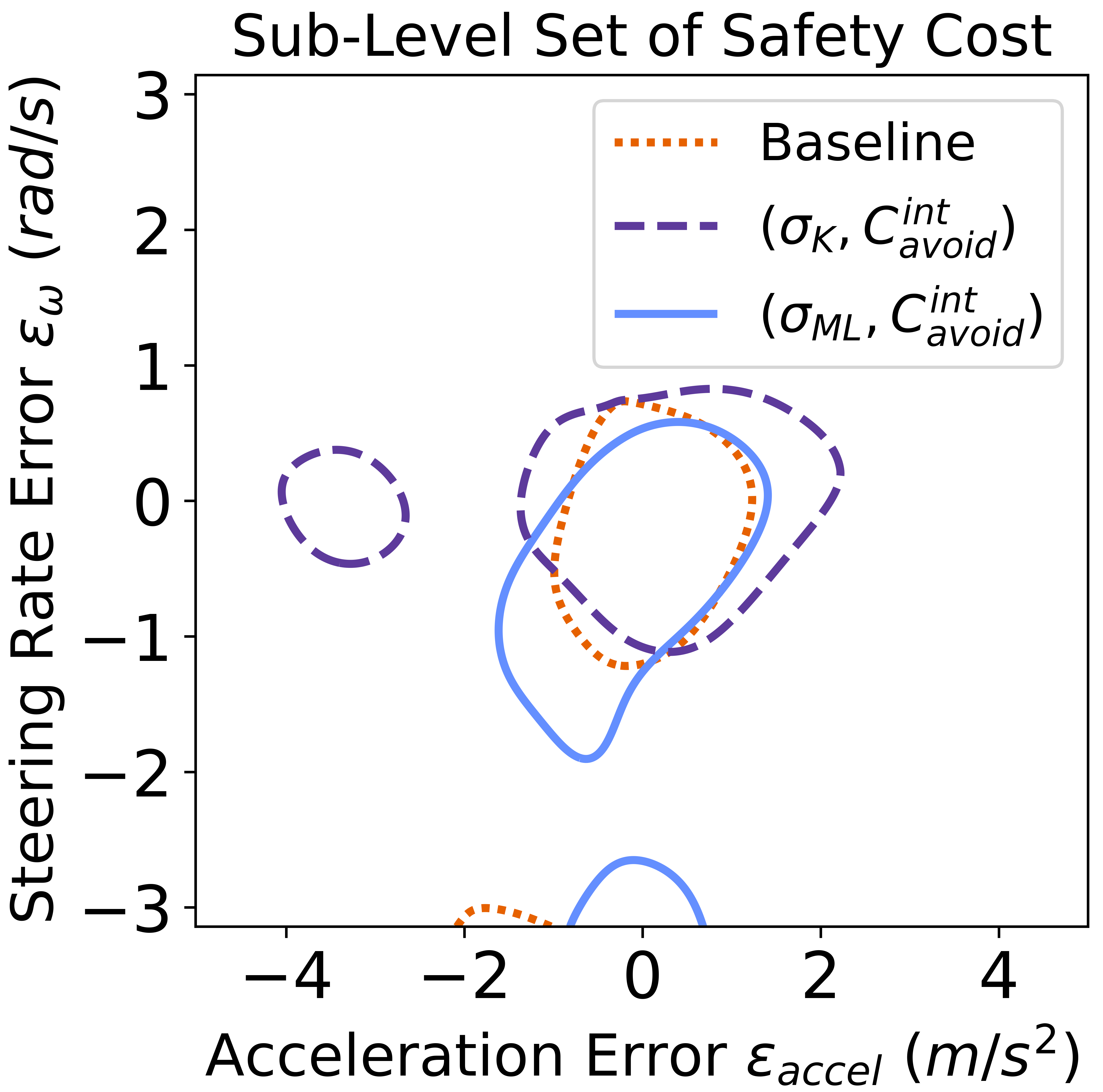}
\caption{(Autonomous Driving System Case Study) We show the sub-level sets of $F(\epsilon)$ for the holistic (top) and safety (bottom) cost functions.
For the holistic cost, $c = 0.05$, and for the safety cost, $c= 0.02$.
We omit sub-level sets of size 0.
Some sub-level sets occupy the whole space.
Part of the sub-level set for the Baseline system for the safety cost is also shown in Figure \ref{fig:cover_figure}.}
\label{fig:nvidia_level_sets}
\end{figure}

\subsubsection{Autonomous Driving System}\label{subsubsec:robustness_av_sys}

For this system, $x$ represents the control history of the target agent for the trajectory predictor.
We define $p_\epsilon(x)$ as representing a perturbation to the acceleration and steering rate of the final, most recent step of the history.
That is, $\epsilon$ is two-dimensional, consisting of $\epsilon_{\text{accel}} \in [-5, 5]$ and $ \epsilon_{\omega} \in [-\pi, \pi]$.
We add $\epsilon$ to the last control action and re-compute the history trajectory according to the agent dynamics. 

The system designs that we will compare will involve different combinations of uncertainty heuristics and uncertainty-aware planner designs: $(\sigma_{ML}, C_{\text{avoid}}^{\text{int}})$, $(\sigma_{ML}, C_{\text{lane}}^{\text{int}})$, $(\sigma_{K}, C_{\text{avoid}}^{\text{int}})$, and $(\sigma_{K}, C_{\text{lane}}^{\text{int}})$.
We will also involve the baseline system in our comparison.
For our cost functions, we use the same holistic and safety cost functions as in Section \ref{subsec:nvidia_analysis_specifications}.
Similar to that section and in our Running Example, we take costs \textit{relative} to that produced by a system with access to the ground truth target agent future.
This is because autonomous driving offers a system a wide variety of possible scenarios in which it is difficult to assess performance using absolute cost. 
For example, a cost may be high because the scenario is inherently difficult to navigate, or inherently low because perhaps all non-ego agents are very far away and do not impact the ego agent.

For this Case Study, we fit the GPs that represent $F(\epsilon)$ over the two-dimensional space of $\epsilon$.
We first take 15 random samples of $\epsilon$ for an initial GP fitting and then use the MILE acquisition function to sample 20 additional times.
The cost threshold $c$ is set to 0.05 for the holistic cost and 0.02 for the safety cost.
We set $\gamma$ from Equation (\ref{eqn:gp_probabilistic_guarantee}) to 0.9.
In Figure \ref{fig:nvidia_level_sets}, we display the obtained level sets.
For each $\epsilon$ sample, we evaluate each system on a subset of the nuScenes dataset that is unseen by Trajectron++ at training time and compute the average cost.
This amounts to 12,452 data points.
Again, each $F(\epsilon)$ evaluation takes on the order of 15 minutes.
The GP priors have zero means and covariances specified by Matern52 kernel functions.
To calculate the sizes of the level sets, we grid the acceleration error axis in increments of 0.02 and the steering rate error axis in increments of 0.015.

In Table \ref{tab:nvidia_robustness_results}, we compare the performance and robustness of the systems.
We find that for both the holistic and safety costs, there exists a design involving one of our uncertainty measures and uncertainty-aware planners that is better than the baseline in terms of both performance and robustness.
As one might expect, using the agent-avoiding planner is best for safety cost performance and robustness.
We also point out the particular efficacy of using the lane-keeping planner in terms of the holistic cost performance and robustness.
This shows the promise of lane-keeping behavior as a strategy in the presence of input errors.
The coupling between the uncertainty measure and the decision-making-under-uncertainty algorithm, however, proves to be very important.
For example, the agent-avoiding planner has a good holistic cost when $\sigma_{ML}$ is used, but a poor cost when $\sigma_{K}$ is used.

We can also more deeply analyze the sub-level sets in Figure \ref{fig:nvidia_level_sets} to yield useful insights about the system.
For instance, the holistic cost level sets tend to have volume when $\epsilon_{\omega} < 0$ but not when $\epsilon_{\omega} > 0$.
We found two reasons for this.
For one, we found that our trajectory predictor Trajectron++ performs better when $\epsilon_{\omega} < 0$ than when $\epsilon_{\omega} > 0$, perhaps due to biases in the training data.
Secondly, when the steering rate error is negative the target agent's prediction is often such that the ego agent's plan is less likely to intersect with it and can more directly reach the goal.

Thus, our method serves to not only help a designer measure a design's robustness, but also produce a better understanding of the system.
It is not only a tool for comparing designs but also for helping to diagnose vulnerabilities and weaknesses in potential future operating conditions.
This can additionally be helpful from a system safety validation standpoint: our analysis provides a designer with the set of operating conditions for which a system performs well in addition to the conditions for which it does not.
Subsequent data-gathering efforts can then be concentrated on the regimes of lower performance.

\begin{table}[] 
\centering 
\caption{Sub-Level Set Sizes and Costs for AV Designs} 
\label{tab:nvidia_robustness_results} 
\hspace*{-0.3cm}
\begin{tabular}{|c|*{4}{c|}}

\hline

System & \multicolumn{2}{c|}{Holistic Cost} & \multicolumn{2}{c|}{Safety Cost} \\

\cline{2-5}

& Size of & Cost & Size of & Cost \\

& Sub-Level& $\times 10^{-2}$ $\downarrow$& Sub-Level& $\times 10^{-2}$ $\downarrow$ \\

& Set $\uparrow$& \hfill & Set $\uparrow$& \hfill \\

\hline

GT Prediction & --- & 0 & --- & 0 \\

\hline

Baseline & 0.092 & 1.44 & 0.036 & 0.46 \\

\hline

$(\sigma_{ML}, C_{\text{avoid}}^{\text{int}})$ & 0.0 & \textbf{-0.76} & 0.036 & 0.38 \\

\hline

$(\sigma_{ML}, C_{\text{lane}}^{\text{int}})$ & \textbf{1.0} & -0.49 & 0.0 & 7.77 \\

\hline

$(\sigma_{K}, C_{\text{avoid}}^{\text{int}})$ & 0.084 & 2.90 & \textbf{0.073}
 & \textbf{-2.63} \\

\hline

$(\sigma_{K}, C_{\text{lane}}^{\text{int}})$ & \textbf{1.0} & -0.19 & 0.0 & 8.28 \\

\hline

\end{tabular}
\end{table}

\section{Conclusion}\label{sec:conclusion}

In this paper, we present both a theoretical analysis framework for understanding module-level uncertainty quantification from a system-level perspective as well as an empirical application of the analyses to two real-world systems.
Our analyses are twofold: (1) We use assume-guarantee contract theory to produce a probabilistic specification on a module's uncertainty measure, and (2) We use sub-level set estimation techniques to produce a system-level robustness metric for comparing uncertainty-aware designs.
Concretely, both techniques allow a designer to select between uncertainty quantification methods and decision-making algorithm designs in a principled manner and to understand why a particular design choice behaves the way it does.
However, at a more conceptual level, we also position our work as providing more understanding of what an uncertainty measure truly means. 
At the computational level, we treat system components as black boxes to enable generality, thus shifting the analysis burden to the \textit{data} and making our computational complexity a function of dataset sizes and simulator runtimes.

We also imagine the potential of our work to contribute to system validation processes and to producing new designs.
Both of our analysis techniques involve variables that are the choice of the designer, and each setting of these variables may lead to the discovery of a different system vulnerability or behavior.
The data-driven nature of our work means that the designer can determine which data points led to failure and perform further data gathering.
Some of the variables and expressions introduced by our formulations could even be subject to optimization so as to produce better designs.
In this way, our work can aid a system designer in producing robust-by-design architectures.

One limitation of our work lies in our analyses' assumption that the system is open-loop and time-invariant.
In future work, we would like to perform a similar analysis for closed-loop systems, in which uncertainty measures may be influenced by states and observations at previous time steps.

\section*{Acknowledgments}
This material is based upon work supported by the DARPA Assured Autonomy Program, the NASA ULI on 
Safe Aviation Autonomy, and the NSF GRFP under Grant No. 2146752. 
Any opinions, findings, and conclusions or recommendations expressed in this material are those of the authors and do not necessarily reflect the views of any aforementioned organizations. 
Additionally, the authors would like to thank Brandon Schwiesow, Jose Medina, Zachary Tane, Blake Edwards, James Paunicka, and the rest of the team at Boeing for their help and contributions.

\bibliographystyle{IEEEtran}
{\small
\bibliography{main}  
}

\section{Biography Section}

\vspace{-33pt}
\begin{IEEEbiographynophoto}{Sampada Deglurkar}
Sampada Deglurkar is a Ph.D. student at the University of California, Berkeley in the Electrical Engineering and Computer Sciences department.
She received her B.S. in Electrical Engineering and Computer Science from UC Berkeley in 2020. 
Her research interests are in safe autonomous systems, decision-making under uncertainty, and the principles of design for learning-enabled systems.
She is a recipient of the National Science Foundation’s Graduate Research Fellowship.
\end{IEEEbiographynophoto}

\vspace{-24pt}

\begin{IEEEbiographynophoto}{Haotian Shen}
Haotian Shen received his B.S. (2023) and M.S. (2024) in Electrical Engineering and Computer Sciences from the University of California at Berkeley. 
His research interests include the safety of learning-in-the-loop autonomous systems, uncertainty quantification and reinforcement learning.
He is currently at LinkedIn.
\end{IEEEbiographynophoto}

\vspace{-24pt}

\begin{IEEEbiographynophoto}{Anish Muthali}
Anish Muthali received his B.S. and M.S. in Electrical Engineering and Computer Sciences from the University of California at Berkeley. 
He is currently at Headlands Technologies.
\end{IEEEbiographynophoto}

\vspace{-24pt}

\begin{IEEEbiographynophoto}{Marco Pavone}
Professor Marco Pavone received the Ph.D. degree in aeronautics and astronautics from the Massachusetts Institute
of Technology, Cambridge, MA, USA, in 2010. 
He is an Associate Professor of Aeronautics and Astronautics with Stanford University, Stanford, CA, USA, and leads autonomous vehicle research at NVIDIA, Santa Clara, CA, USA. 
His research interests include the development of methodologies for the analysis, design, and control of autonomous systems, with an emphasis on self-driving cars, autonomous aerospace vehicles, and future mobility systems.
\end{IEEEbiographynophoto}

\vspace{-24pt}

\begin{IEEEbiographynophoto}{Dragos Margineantu}
Dr. Dragos Margineantu is a Boeing Senior Technical Fellow
and Artificial Intelligence (AI) Chief Technologist who is the technical lead of AI research and engineering in Boeing.
His interests include computational methods for robust
systems, autonomous commercial flight, anomaly and surprise detection and handling, reasoning under uncertainty,
validation and testing of decision systems, cost-sensitive, active, ensemble learning, and inverse reinforcement learning.
Dragos was one of the pioneers in research on ensemble
learning and cost-sensitive learning and on statistical testing of learned models.
At Boeing, he developed assurance methods for decision
systems, machine learning based solutions for autonomous
flight, airplane maintenance, airplane performance,
surveillance, and security.

\end{IEEEbiographynophoto}

\vspace{-24pt}

\begin{IEEEbiographynophoto}{Peter Karkus}
Dr. Peter Karkus is a Senior Research Scientist at NVIDIA, Autonomous Vehicles Research Group. 
His interests include autonomous driving, robotics and machine learning, with particular focus on end-to-end architectures that combines scalable learning with structure and reasoning. 
He has received a Ph.D degree from the National University of Singapore in 2021. 
He previously held research appointments at CMU, MIT, Google, and DeepMind. 
\end{IEEEbiographynophoto}

\vspace{-24pt}

\begin{IEEEbiographynophoto}{Boris Ivanovic}
Dr. Boris Ivanovic is a Senior Research Scientist and Manager in NVIDIA’s Autonomous Vehicle Research Group, developing novel end-to-end AV architectures, policy training strategies, sensor and traffic simulation, AI safety, and the integration of foundation models in AV development. 
Prior to joining NVIDIA, he received his Ph.D. in Aeronautics and Astronautics and an M.S. in Computer Science, both from Stanford University.
\end{IEEEbiographynophoto}

\vspace{-24pt}

\begin{IEEEbiographynophoto}{Claire J. Tomlin}
Professor Claire Tomlin holds the James and Katherine Lau Chair in Engineering at UC Berkeley.
Her research interests include hybrid systems, distributed and decentralized optimization, and control theory, with an emphasis on applications, unmanned aerial vehicles, air traffic control and modeling of biological processes. 
She taught at Stanford University from 1998 to 2007 where she was a director of the Hybrid Systems Laboratory and held joint positions in the Department of Aeronautics and Astronautics and the Department of Electrical Engineering. 
She was awarded a MacArthur Genius grant in 2006 and the IEEE Transportation Technologies Award in 2017 ``for contributions to air transportation systems, focusing on collision avoidance protocol design and avionics safety verification''.
\end{IEEEbiographynophoto}

\vfill

\end{document}